\documentclass{aastex}
\usepackage{spr-astr-addons}
\usepackage{url}

\RequirePackage{color}

\newcommand{\HII}{H {\footnotesize{II}} }
\newcommand{\kms}{{\rm km~s}^{-1}}

\begin{document}

\title{Gas structure and dynamics towards bipolar infrared bubble}
\slugcomment{Not to appear in Nonlearned J., 45.}
\shorttitle{Gas structure and dynamics towards bipolar infrared bubble}
\shortauthors{Xu et al.}

\author{Jin-Long Xu} 
\affil{National Astronomical Observatories, Chinese Academy of Sciences, Beijing 100012, China}
\and 
\author{Naiping Yu}  
\affil{National Astronomical Observatories, Chinese Academy of Sciences, Beijing 100012, China}
\and 
\author{Chuan-Peng Zhang}
\affil{National Astronomical Observatories, Chinese Academy of Sciences, Beijing 100012, China}
\and 
\author{Xiao-Lan Liu}
\affil{National Astronomical Observatories, Chinese Academy of Sciences, Beijing 100012, China}


\begin{abstract}
We present multi-wavelength analysis for four bipolar bubbles (G045.386-0.726, G049.998-0.125, G050.489+0.993, and G051.610-0.357) to probe the structure and dynamics of their surrounding gas. The $^{12}$CO $J$=1-0, $^{13}$CO $J$=1-0 and C$^{18}$O $J$=1-0 observations are made with the Purple Mountain Observation (PMO) 13.7 m radio telescope. For the four bipolar bubbles, the bright 8.0 $\mu$m emission shows the bipolar structure. Each bipolar bubble is associated with an \HII region. From CO observations we find that G045.386-0.726 is composed of two bubbles with different distances, not a bipolar bubble.  Each of G049.998-0.125 and G051.610-0.357 is associated with a filament. The filaments in CO emission divide G049.998-0.125 and G051.610-0.357 into two lobes. We suggest that the exciting stars of both G049.998-0.125 and G051.610-0.357  form in a sheet-like structure clouds. Furthermore, G050.489+0.993 is associated with a clump, which shows a triangle-like shape with a steep integrated intensity gradient towards the two lobes of G050.489+0.993. We suggest that the two lobes of G050.489+0.993 have simultaneously expanded into the clump. 
\end{abstract}

\keywords{\HII regions --- ISM: bubbles --- ISM: clouds}

\label{sec:intro}
\section{Introduction}
Infrared bubbles are the bright 8.0 $\mu$m emission surrounding O and early-B stars. The bright 8 $\mu$m emission is attributed to polycyclic aromatic hydrocarbons \citep[PAHs,][]{Leger1984}. The PAHs molecules can be destroyed inside the ionized gas, but are excited in the photodissociation region (PDR) by the UV radiation within \HII region \citep{Pomares2009}.  \HII regions are ionized by massive stars. The UV radiation of an \HII region can excite its surrounding gas to create a bright mid-infrared bubble. So the bubbles are usually found in or near massive star-forming regions \citep[see e.g.,][]{Deharveng2003,Xu2014a,Xu2014b}.  If bubble forms in  a sheet-like molecular cloud, whose thickness is not greater than the bubble size, we will see  a 2D ring bubble. And when the bubble size is less than the thickness of the molecular cloud, a 3D spherical-structure bubble will form. Hence,  the environment of bubble formed may determine their morphology. 

\begin{table*}[t]
\small
\tabcolsep 1.3mm\caption{Physical parameters of \HII regions and molecular gas associated with four bipolar bubbles\label{bubble}}
\begin{tabular}{@{}lcccccccccccc@{}}
\tableline
Name &  R.A. & Decl.  & S & $V_{\rm H}$ & Distance & $N_{\rm Ly}^{c}$ & Type$^{d}$ & $N(\rm H_{2}$) &  $M(\rm H_{2}$)\\
  & (J2000)  &  (J2000)   &(mJy)& km $\rm s^{-1}$ &(kpc) & (photons~$\rm s^{-1}$) & & (cm~$^{-2}$)& M$_{\odot}$\\
\tableline
G045.386-0.726 &  19$^{\rm h}$17$^{\rm m}$00.5$^{\rm s}$ &  10$^{\rm \circ}$44$^{\rm \prime}$33$^{\rm \prime\prime}$ &   610&52.5 &  8.0 &  7.4$\times$10$^{48}$  & O6.5$\sim$O6 & -- & --\\
G049.998-0.125 & 19$^{\rm h}$23$^{\rm m}$46.1$^{\rm s}$ &  15$^{\rm \circ}$04$^{\rm \prime}$54$^{\rm \prime\prime}$ &   410& 38.5/73.2 & 4.7$^{b}$  & 1.7$\times$10$^{48}$  &O8.5$\sim$O8 &5.6$\times$10$^{21}$ & 8.3$\times$10$^{3}$ \\
G050.489+0.993& 19$^{\rm h}$20$^{\rm m}$38.2$^{\rm s}$ &  16$^{\rm \circ}$02$^{\rm \prime}$29$^{\rm \prime\prime}$ &   140$^{a}$& 57.1&  5.4  &  6.4$\times$10$^{47}$ & $<$O9.5  &3.1$\times$10$^{21}$ & 4.4$\times$10$^{3}$\\
G051.610-0.357& 19$^{\rm h}$27$^{\rm m}$48.0$^{\rm s}$ &  16$^{\rm \circ}$23$^{\rm \prime}$25$^{\rm \prime\prime}$ &   270& 41.1 &  5.3  &   1.4$\times$10$^{48}$ & O8.5$\sim$O8  &3.6$\times$10$^{21}$  & 3.7$\times$10$^{3}$\\
\tableline
\end{tabular}
\vspace{-5mm}
\tablenotetext{a}{The integrated flux density is obtained from the NVSS 1.4 GHz radio continuum.}
\tablenotetext{b}{ The distance is measured from this paper.}
\tablenotetext{c}{The ionizing luminosity $N_{\rm Ly}$ is computed by \citet{Mezger1974}.}
\tablenotetext{d}{\citet{Martins2005}.}
\end{table*}

Bipolar bubble has a simple  morphology, so it is easy to locate the ionized and neutral components in space. So far very few studies describe the formation and  evolution of the bipolar bubble. \citet{Deharveng2010} used the ATLASGAL survey at 870 $\mu$m to study 102 bubbles detected by spitzer-GLIMPSE, 86$\%$ of which enclose classical \HII region ionized by O and B stars. Only bubbles N39 and N52 show a bipolar morphology \citep{Churchwell06}, which are associated with a filament. They considered that the ionized stars of their two bipolar bubbles formed in a massive and rather flat molecular cloud.  \citet{Deharveng2015} identified and studied two bipolar bubbles. \HII regions G319.88+00.79 and G010.32-00.15 are located at the centre of these two  bipolar bubbles, respectively. Especially for \HII regions G319.88+00.79, there is a filament that will divide the bipolar bubble into  two ionized lobes. They concluded that the exciting stars of each bipolar \HII region are formed in a sheet-like structure clouds. The sheet observed edge-on appears as a filament. According to \citet{Deharveng2010,Deharveng2015}, the bipolar bubble forms when ionization front breaks through the two opposite faces of the sheet-like cloud simultaneously. Thus, the observation of bipolar bubbles can  also provide important information on the structure of their surrounding cloud.

\begin{figure*}[t]
   \centering
  \includegraphics[width=6.7cm, angle=0]{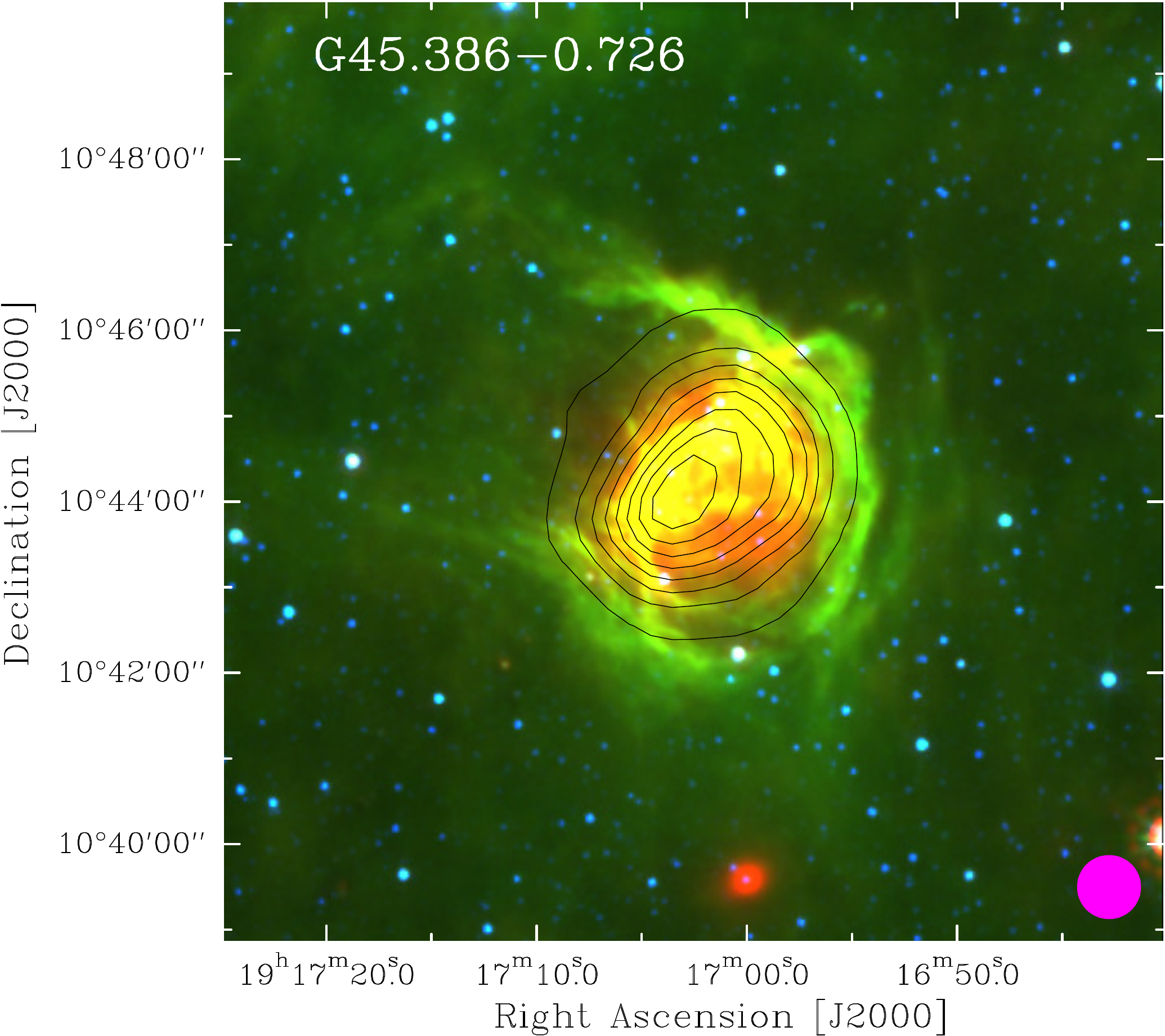}
  \includegraphics[width=6.7cm, angle=0]{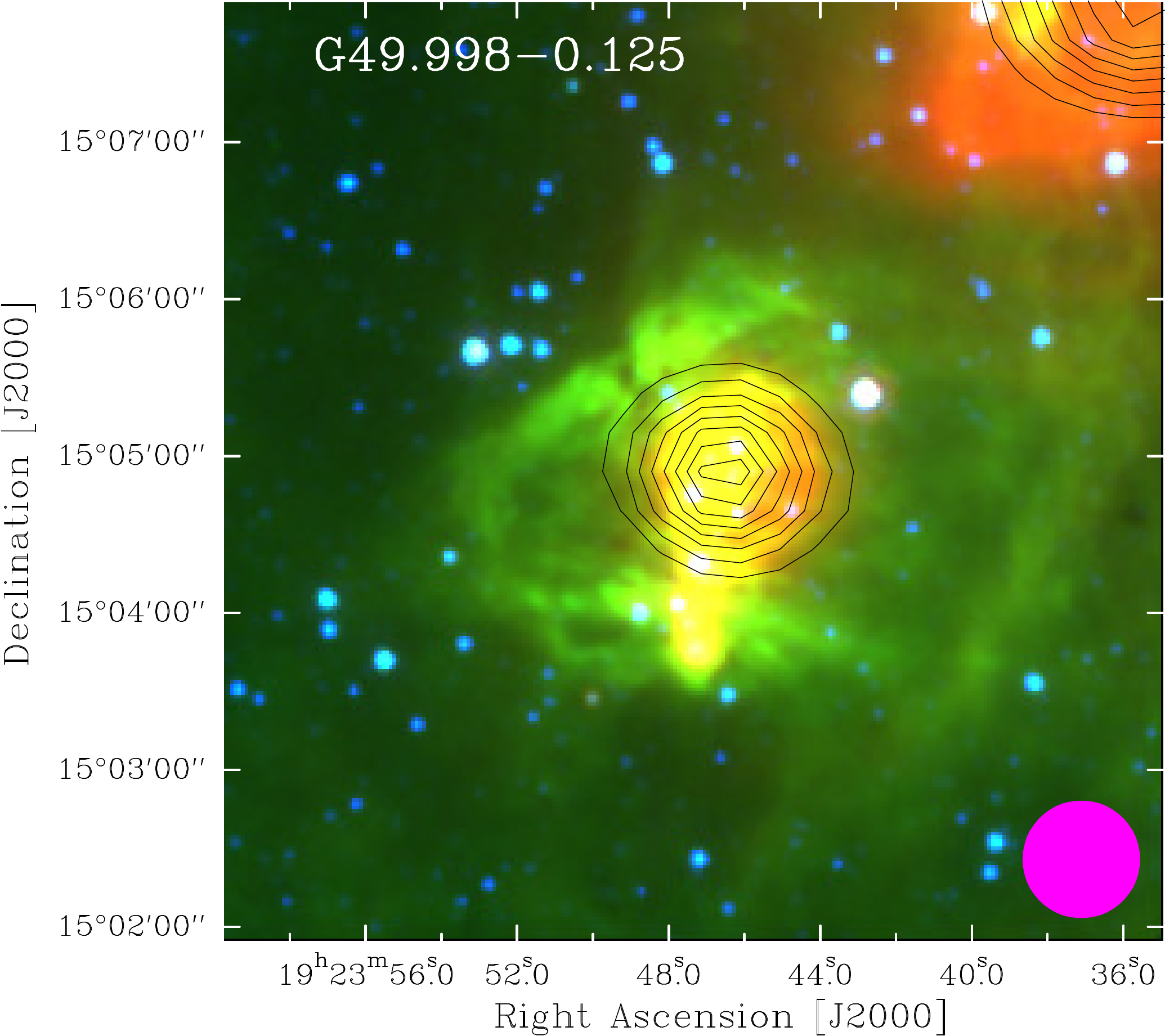}
  \includegraphics[width=6.7cm, angle=0]{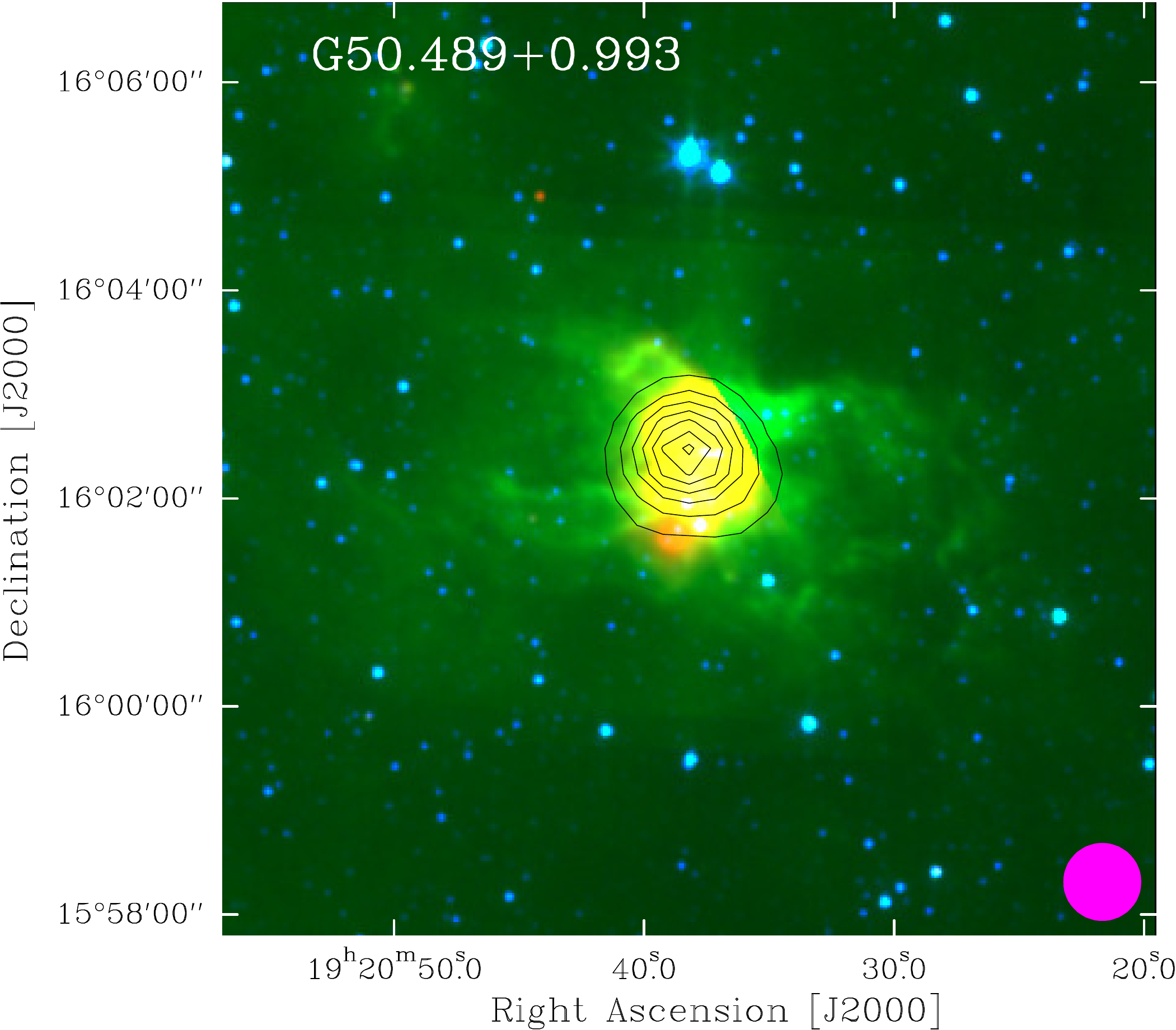}
  \includegraphics[width=6.7cm, angle=0]{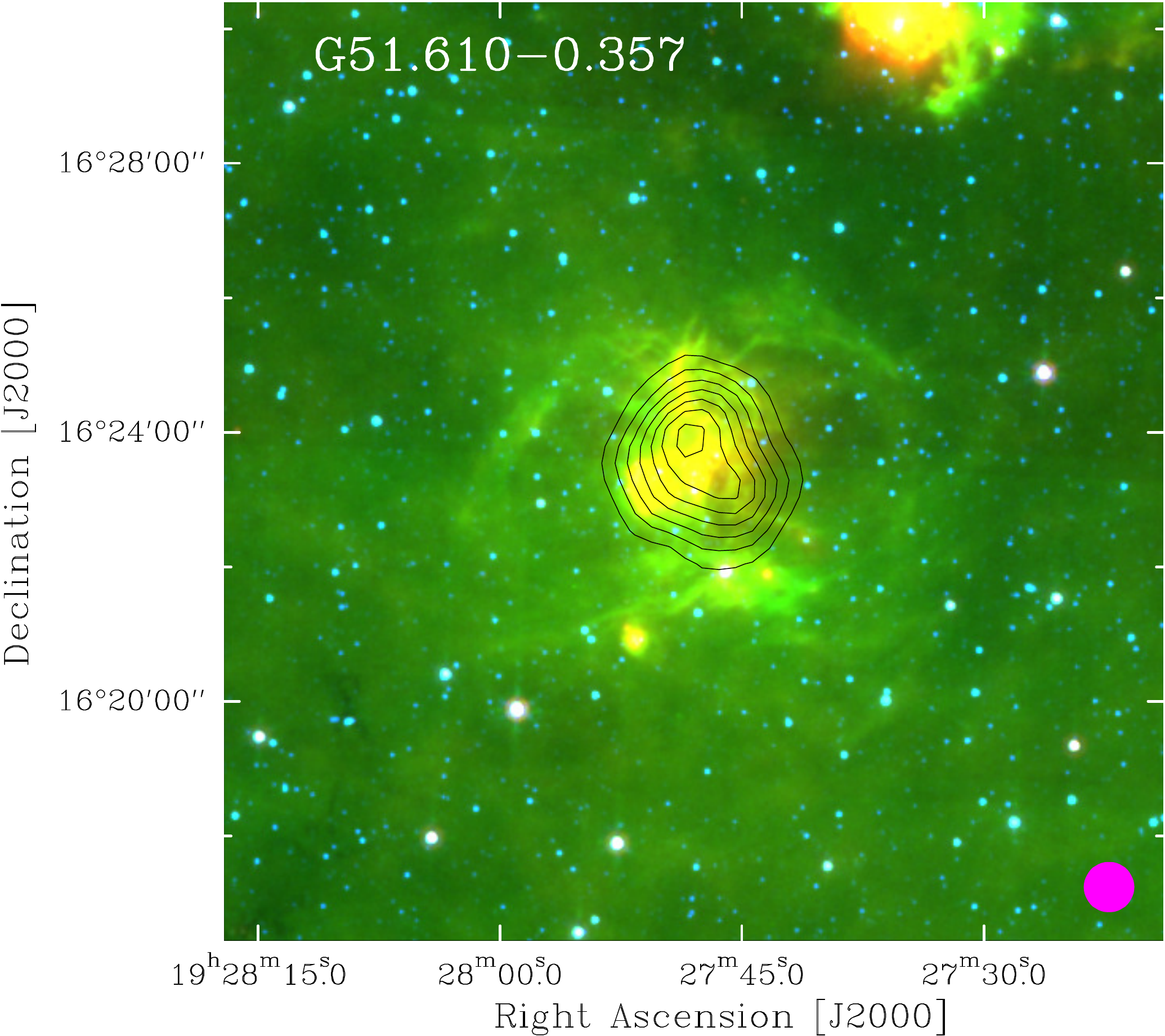}
\vspace{-4mm}
   \caption{1.4 GHz radio continuum contours in black colour, overlaid on the three color image of bipolar bubble composed from the Spitzer 4.5 $\mu$m (in blue), 8.0 $\mu$m (in green), and 24 $\mu$m (in red) bands. G045.386-0.726: The black contour levels are 3.5 (5$\sigma$), 14.0, 24.5, 35.0, 45.5, 56.0, 66.5, and 77.0 mJy beam$^{-1}$; G049.998-0.125: The black contour levels are 4.5 (5$\sigma$), 7.1, 9.8, 12.5, 15.1, 17.8, 20.5, and 23.2 mJy beam$^{-1}$; G050.489+0.993: The black contour levels are 3.1 (5$\sigma$), 6.1, 9.1, 12.2, 15.3, 18.3,  and 21.4 mJy beam$^{-1}$; G051.610-0.357: The black contour levels are 2.8 (5$\sigma$), 4.5, 6.2, 7.8, 9.5, 11.2, and 12.9 mJy beam$^{-1}$.} 
   \label{Fig1}
   \end{figure*}

According to the Green Bank Telescope (GBT) and Arecibo \HII region discovery surveys, \citet{Anderson2011} and \citet{Bania2012} identified eleven bipolar bubbles. Considering the angular resolution of our used telescope (Purple Mountain Observation 13.7 m radio telescope), we only selected four bipolar bubbles, whose parameters are lised in Table \ref{bubble}, including the coordinates, the integrated flux densities, the hydrogen radio recombination line (RRL) velocities ($V_{\rm H}$), and the distances. In this paper, we performed a multi-wavelength study to the four bipolar bubbles to investigate the gas structure around \HII regions. Our observations and data reduction are described in Sect.\ref{sect:Obs}, and the results are presented in Sect.\ref{sect:results}. In Sect.\ref{sect:discu}, we  discuss the gas structure around the four bipolar bubbles, while our conclusions are summarized in Sect.\ref{sect:summary}.

\section{Observations}
\label{sect:Obs}
\subsection{Purple Mountain Data}
We made the mapping observations for four bipolar bubbles and their adjacent regions in the transitions of $^{12}$CO $J$=1-0, $^{13}$CO $J$=1-0 and C$^{18}$O $J$=1-0 lines using the Purple Mountain Observation (PMO) 13.7 m radio telescope at De Ling Ha in the west of China at an altitude of 3200 meters, during May 2015. The 3$\times$3 beam array receiver system in single-sideband (SSB) mode was used as front end. The back end is a fast Fourier transform spectrometer (FFTS) of 16384 channels with a bandwidth of 1 GHz, corresponding to a velocity resolution of 0.16 km s$^{-1}$ for $^{12}$CO $J$=1-0, and 0.17 km s$^{-1}$ for $^{13}$CO $J$=1-0 and C$^{18}$O $J$=1-0. $^{12}$CO $J$=1-0 was observed at upper sideband, while $^{13}$CO $J$=1-0 and C$^{18}$O $J$=1-0 were observed simultaneously at lower sideband.  The half-power beam width (HPBW) was 53$^{\prime\prime}$ at 115 GHz and the main beam efficiency was 0.5. The pointing accuracy of the telescope was better than 5$^{\prime\prime}$, which was derived from continuum observations of planets (Venus, Jupiter, and Saturn). The source W51D (19.2 K) was observed once per hour as flux calibrator. Mapping observations used the on-the-fly mode. The standard chopper wheel calibration technique is used to measure antenna temperature $T_{\rm A} ^{\ast}$ corrected for atmospheric absorption. The final data was recorded in brightness temperature scale of $T_{\rm mb}$ (K). The data were reduced using the GILDAS/CLASS \footnote{http://www.iram.fr/IRAMFR/GILDAS/} package.

\subsection{Archival Data }
The 1.4 GHz radio continuum emission data were obtained from the
NRAO VLA Sky Survey (NVSS), with a noise of about 0.45 mJy/beam and a resolution of 45$^{\prime\prime}$ \citep{Condon1998}; We extracted 870 $\mu$m data from the ATLASGAL survey \citep{Schuller2009}. The survey was carried out with the Large APEX Bolometer Camera observing at 870 $\mu$m (345 GHz).
The APEX telescope has a full width at half-maximum (FWHM) beam size of  19$^{\prime\prime}$ at this frequency; 
We also utilized mid-infrared data in two bands (4.5 and 8.0 $\mu$m) from the Spitzer GLIMPSE survey  and mid-infrared data in 24 $\mu$m band from the MIPSGAL survey \citep{Benjamin2003,Rieke2004}.
The resolutions at  4.5 $\mu$m and 8.0 $\mu$m are $<$ $2.0^{\prime\prime}$, while the MIPSGAL resolution at 24 $\mu$m is 6$^{\prime\prime}$; We also retrieved the K (2.17 $\mu$m) band images from Two Micron All Sky Survey (2MASS)\citep{Skrutskie2006}.

\section{Results}
\label{sect:results}
\subsection{Infrared and Radio Continuum Images}
Figure \ref{Fig1} shows composite three-color images of the four bipolar bubbles. The three infrared bands are the Spitzer 4.5 $\mu$m (in blue), 8.0 $\mu$m (in green), and 24 $\mu$m (in red).  In Fig. \ref{Fig1}, the PAH emission,  traced by the bright 8 $\mu$m, delineates the structure for each bipolar bubble. Two lobes of  both G050.489+0.993 and  G051.610-0.357 are symmetrical,  and thus they are the standard bipolar bubble. For G049.998-0.125, the left lobe is smaller than the right lobel. Compared with the other three bipolar bubble, G045.386-0.726 shows a complex sructure. The 24 $\mu$m  emission traces heated small dust grains \citep{Watson2008}. The hot dust emission in Fig. \ref{Fig1} displays a dense structure between two lobes for each bipolar bubble.  The NVSS 1.4 GHz radio continuum emission, which can be used to trace ionized gas, is also overlaid in Fig. \ref{Fig1}  in black contours. In Fig. \ref{Fig1},  the ionized gas emission are spatially coincident with the hot dust emission observed at the 24 $\mu$m (red color). For each bipolar bubble, both the ionized gas and hot dust emission show a single structure, suggesting that each bipolar bubble is created by  an \HII region, not adjacent bubbles surrounding distince \HII regions.

To estimate the ionized star types of these \HII regions, we will calculate the ionizing luminosity. Assuming the radio continuum emission is optically thin, the ionizing luminosity $N_{\rm Ly}$ is computed by \citet{Mezger1974}
\begin{equation}
 \mathit{N_{\rm Ly}}=4.76\times10^{48}(\frac{\nu}{\rm GHz})^{0.1}(\frac{T_{e}}{\rm K})^{-0.45}(\frac{S_{\nu}}{\rm Jy})(\frac{D}{\rm kpc})^{2}\rm ~s^{-1},
\end{equation}
Where $\nu$ is the frequency, $T_{e}$ is the effective electron temperature, $S_{\nu}$ is the observed specific flux density, and $D$ is the distance to \HII region. \citet{Anderson2011}  measured the flux densities of  \HII regions G045.386-0.726, G049.998-0.125, and  G051.610-0.357 at 9 GHz. For the \HII region related to G050.489+0.993, we measure its flux density  at 1.4 GHz. Moreover, we adopt an effective electron temperature of 10$^{4}$ K.   \citet{Inoue2001} suggested that only half of Lyman continuum photons from  the central source in a Galactic  \HII region ionizes neutral hydrogen, the remainder being absorbed by dust grains within the ionized region. Finally,  we obtain the ionizing luminosity$N_{\rm Ly}$. According to \citet{Martins2005},  the spectral type of the ionizing star for each \HII region is derived,  which are lised in Table \ref{bubble}.

\begin{figure*}[]
\centering
\includegraphics[width=7.4cm, angle=0]{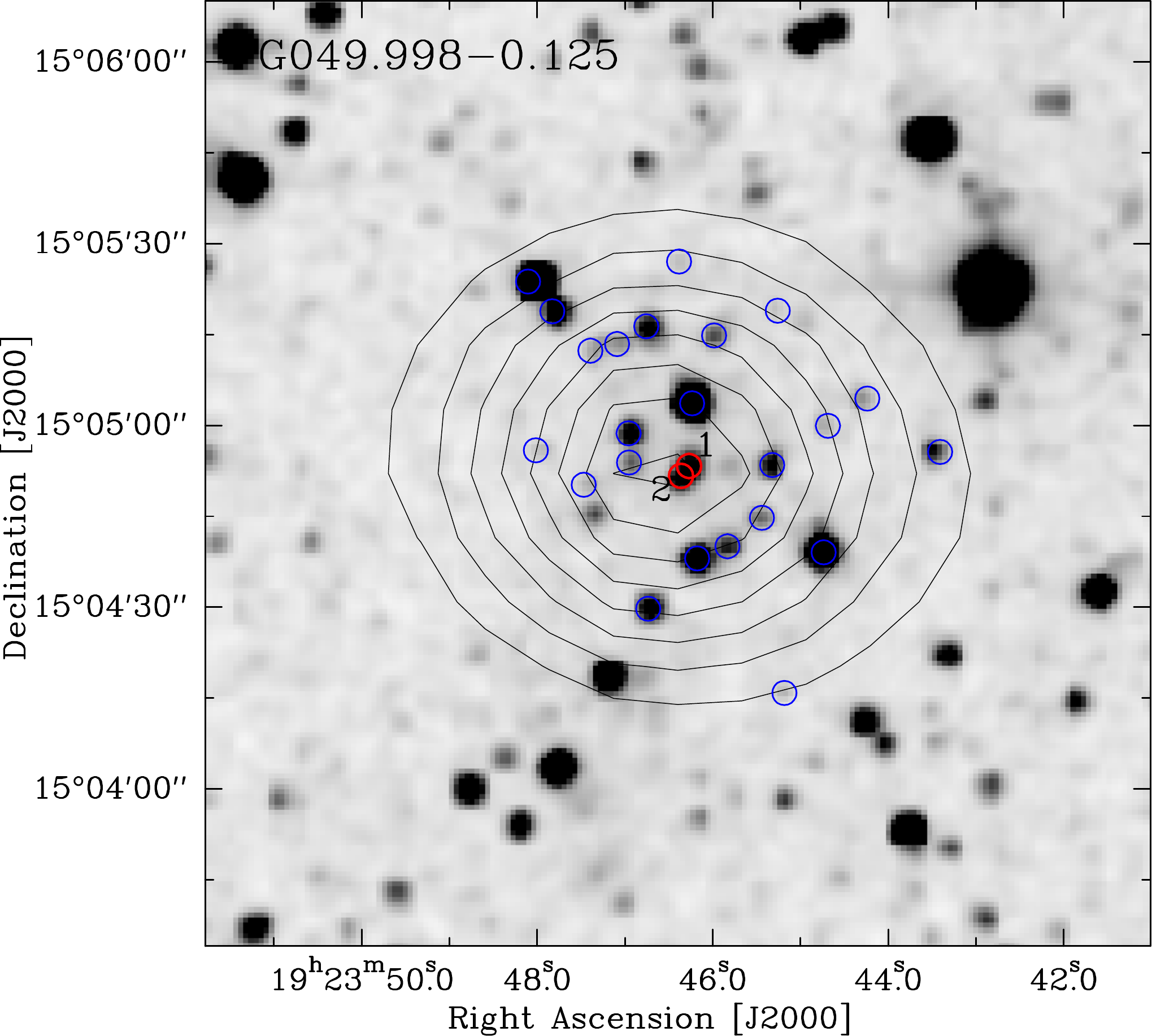}
\includegraphics[width=6.7cm, angle=0]{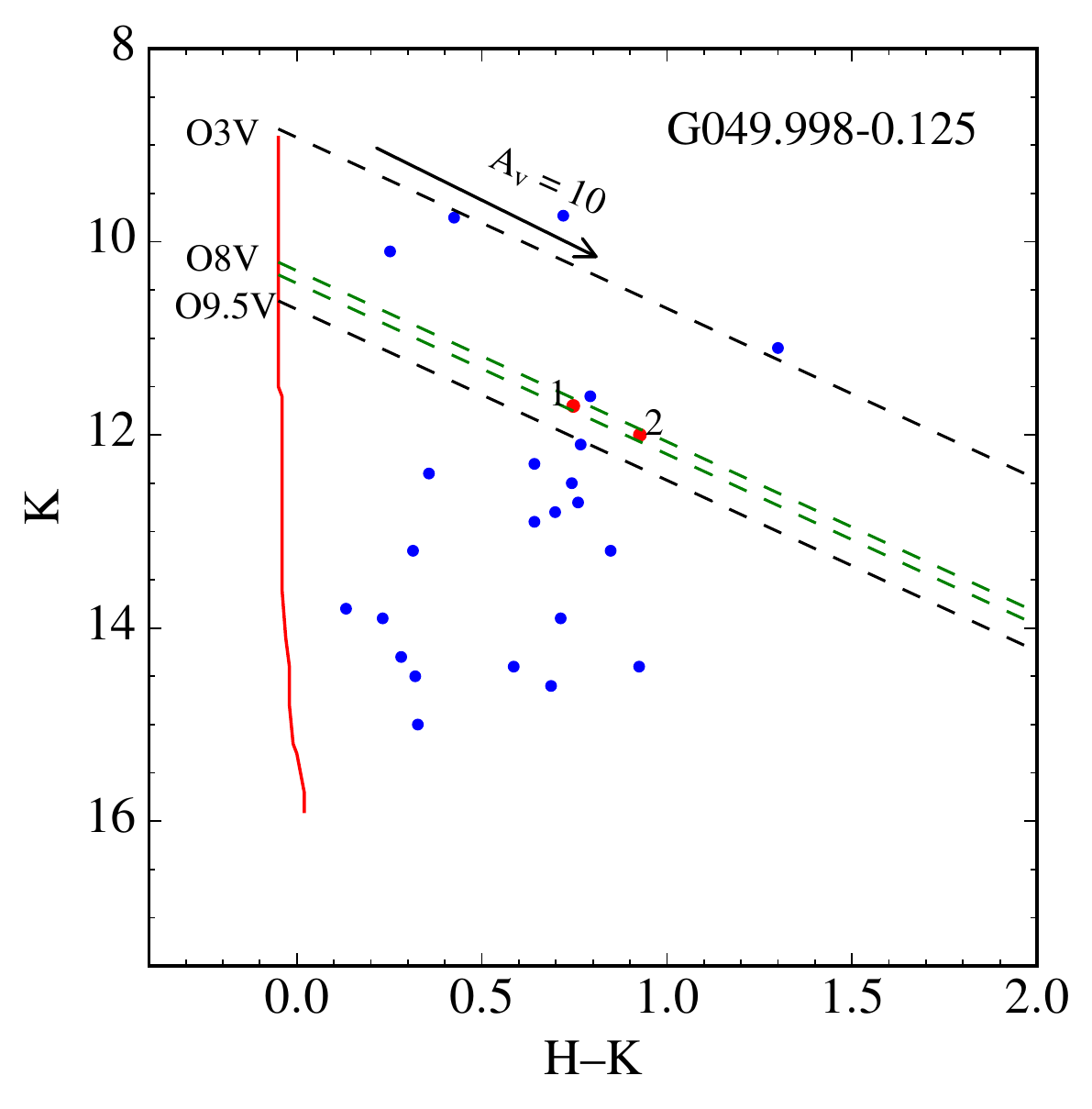}
\includegraphics[width=7.4cm, angle=0]{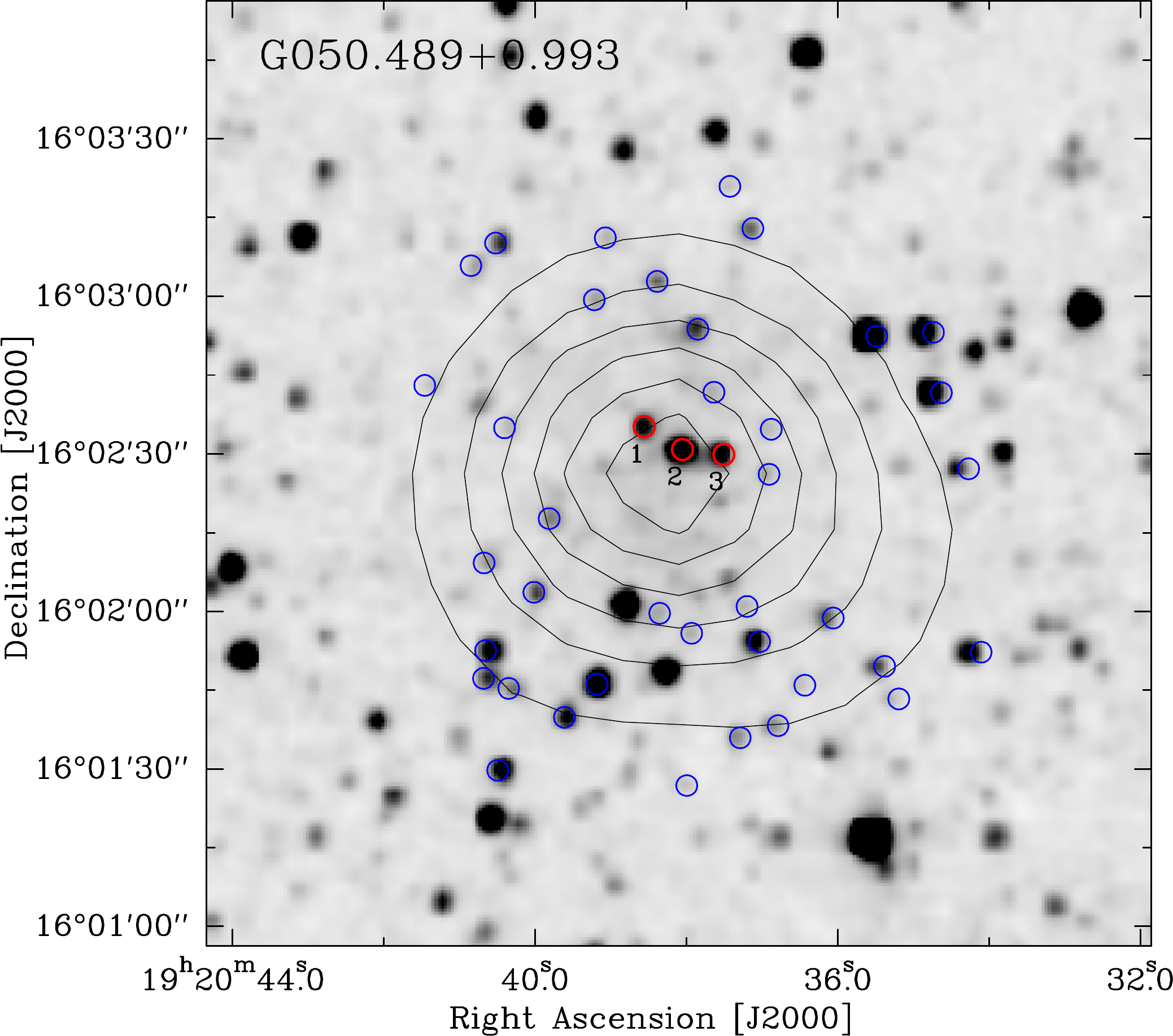}
\includegraphics[width=6.7cm, angle=0]{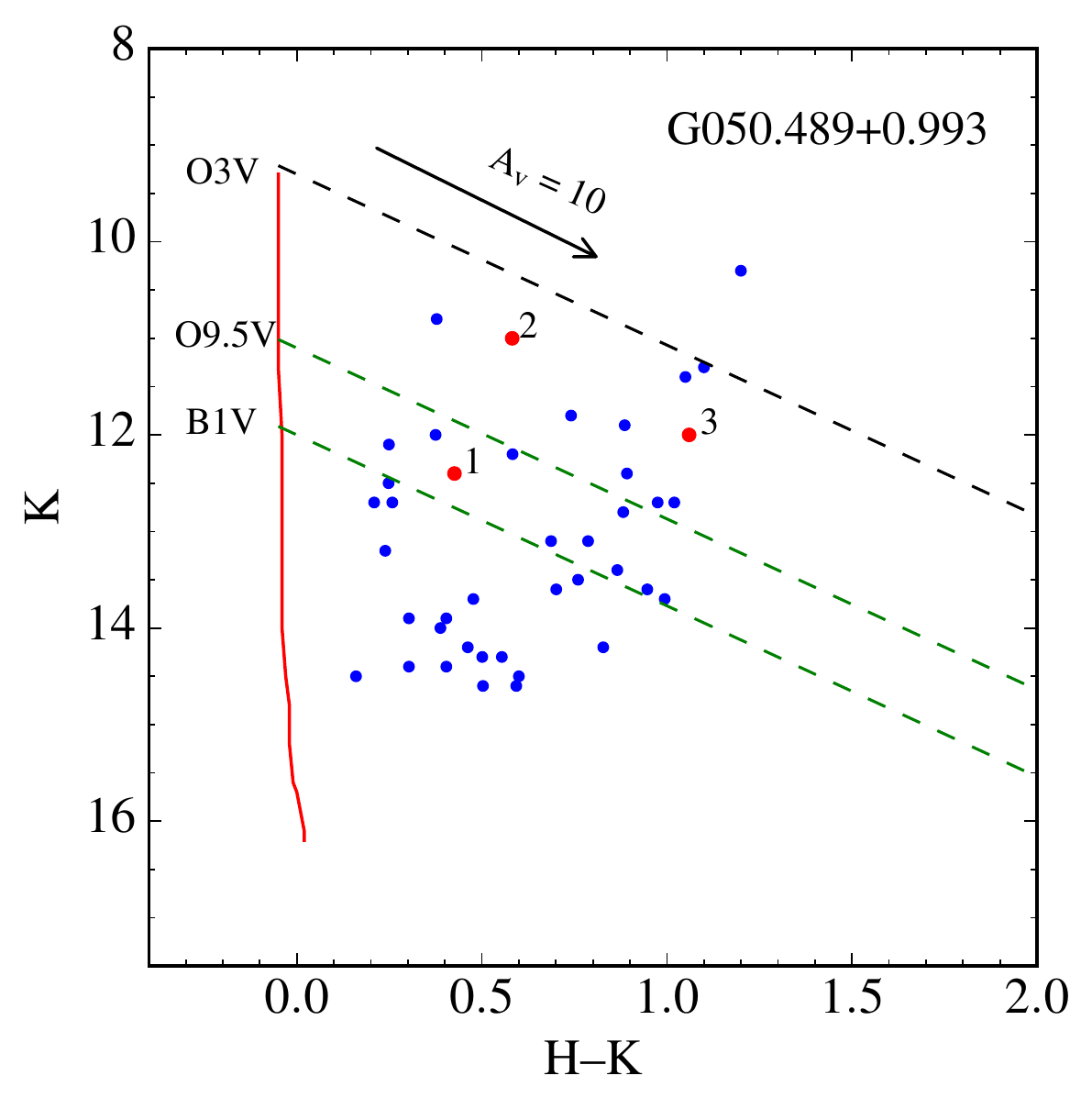}
\includegraphics[width=7.4cm, angle=0]{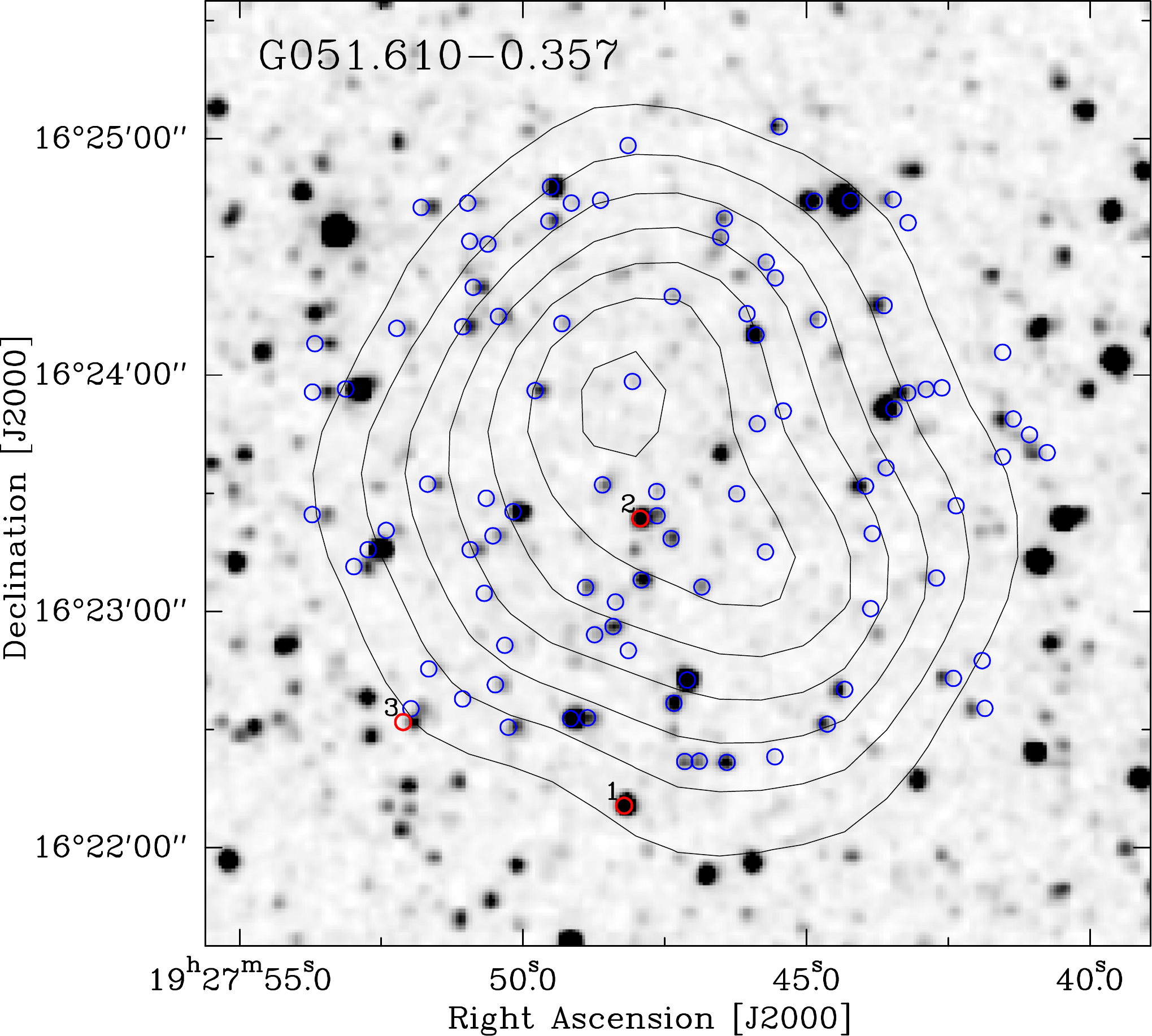}
\includegraphics[width=6.7cm, angle=0]{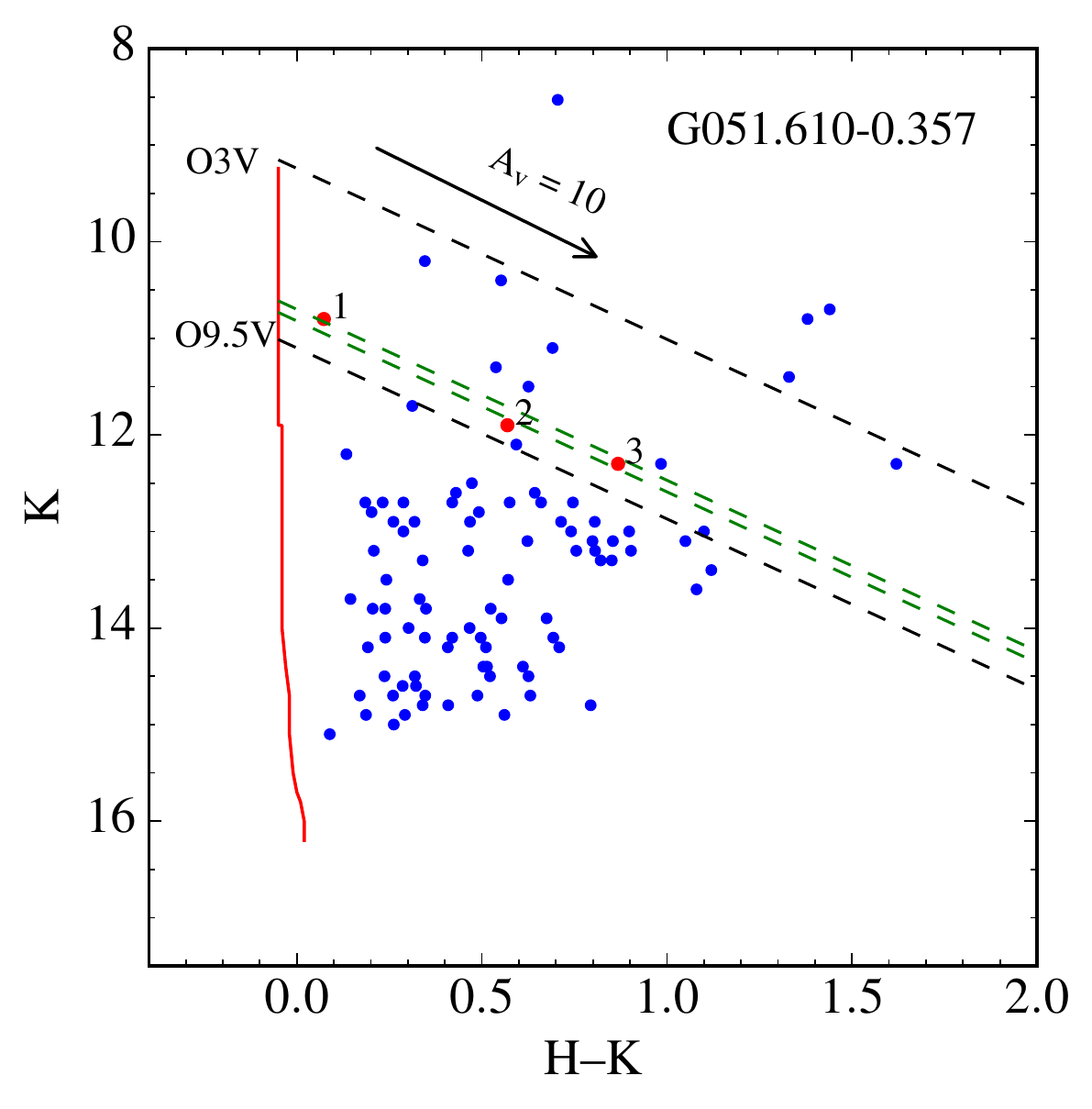}
 \vspace{-4mm}
\caption{Left panels: Radio continuum emission at 1.4 GHz superimposed on the K frame of each \HII region. The blue circles represent the selected near-infrared sources, while the red circles may indicate the ionized stars. Left panels: K versus H--K diagram for sources detected in
all the three 2MASS bands. The main sequence is also drawn for different distances (red lines). The slanting dashed lines trace the reddening vectors for each spectral type (extinction law of \citet{Rieke1985}). While the slanting green dashed lines represent the positions of O8.5V and O8V, respectively. } 
   \label{Fig2}
   \end{figure*}

Figure \ref{Fig2} (left panels) only shows the K frames overlaid with 1.4 GHz continuum emission for three bipolar bubbles (G049.998-0.125, G050.489+0.993, and G051.610-0.357). G045.386-0.726 is not a real bipolar bubble (see Sec.3.2.1). To further look for the ionized star types of these \HII regions,
we use the 2MASS All-Sky Point source database in the near-infrared J(1.25 $\mu$m), H(1.65 $\mu$m) and K(2.17 $\mu$m) bands, with a signal-to-noise ratio (S/N) greater than 10.  Because the 1.4 GHz radio continuum emission can trace the ionized gas of each bipolar bubble,  the ionized stars are only located within the 1.4 GHz radio continuum emission regions. Using the database, we select near-infrared sources, and distribute these sources over the K band images. Figure \ref{Fig2} (right panels) shows the (H) versus (H--K) color-color (CC) diagrams.  
Table \ref{bubble} gives that the ionized star types of both G049.998-0.125 and G051.610-0.357 are between O8.5V and O8V.  From Fig. \ref{Fig2}, we see that two near-infrared sources are located between O8.5V and O8V for G049.998-0.125, and three near-infrared sources near and between O8.5V and O8V for G051.610-0.357. In Fig. \ref{Fig2} (left panels), the red circles mark the positions of these selected near-infrared sources.  Generally, the ionized stars are located near or on the peak position of the ionized gas emission. From  Fig. \ref{Fig2} (left panels),  we suggest that \#1 and \#2 may be the ionized stars of G049.998-0.125, and \#2 is likely to be the ionized star of G051.610-0.357. For G050.489+0.993, the type of its ionized star is greater than O9.5V. Several near-infrared sources are below the 09.5V line, as shown in  Fig. \ref{Fig2} (right panels) for  G050.489+0.99, but we see that three bright near-infrared sources are located near the peak postion of the ionized gas. Comparing  the CC diagram, \#1 may be the ionized stars of G050.489+0.993.
  
  \begin{figure*}[t]
   \centering
  \includegraphics[width=6.7cm, angle=0]{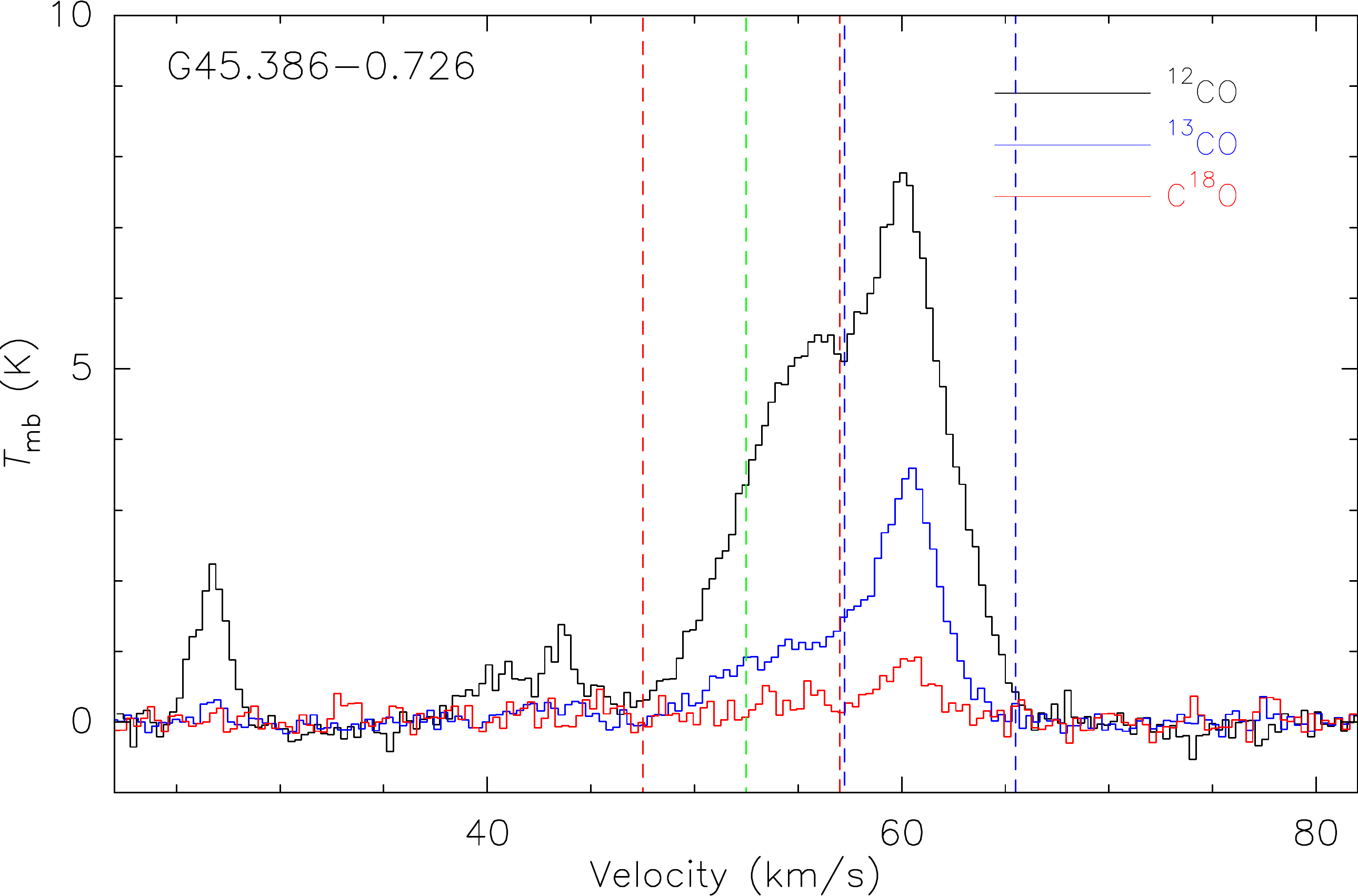}
  \includegraphics[width=6.7cm, angle=0]{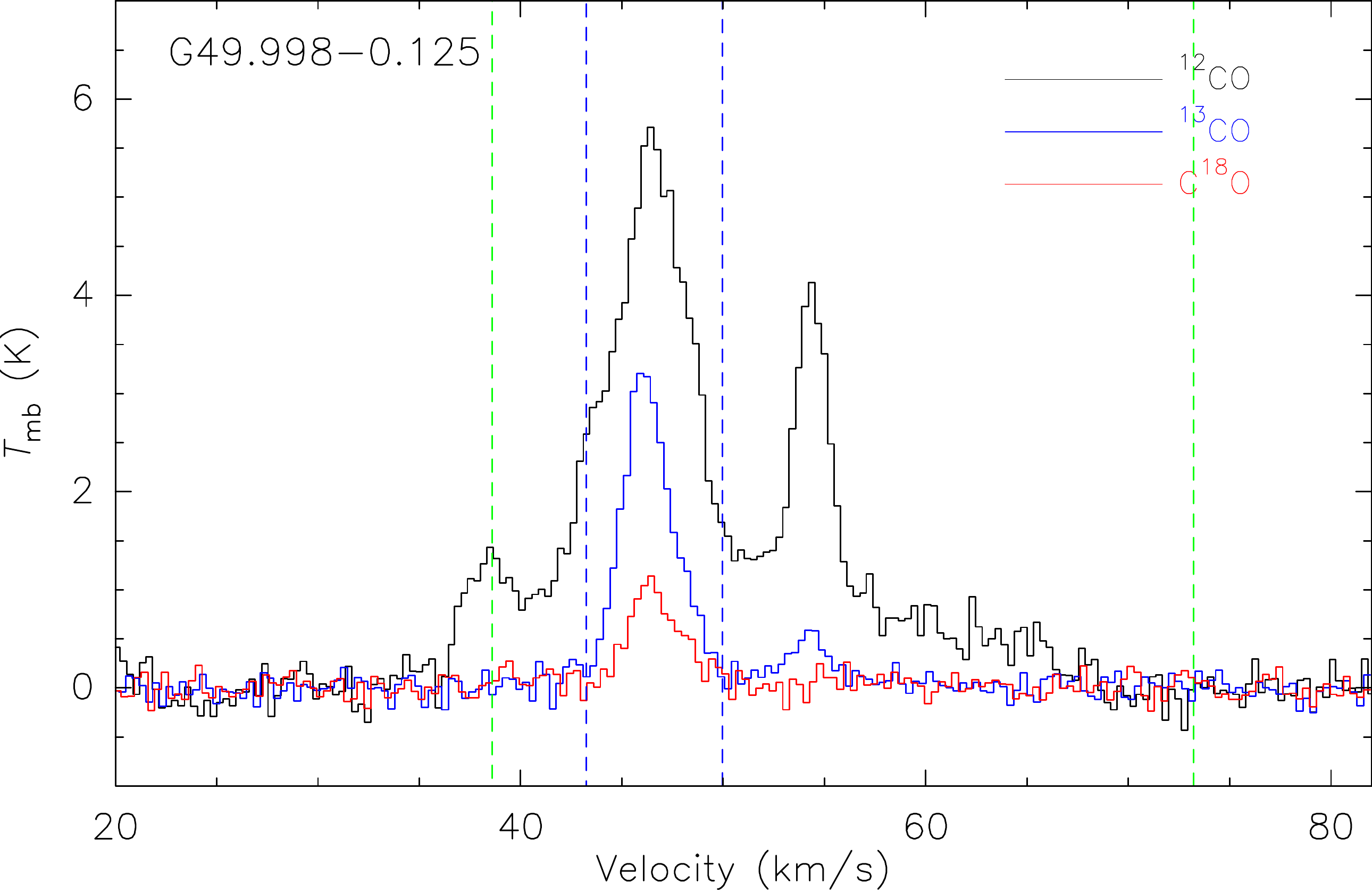}
  \includegraphics[width=6.7cm, angle=0]{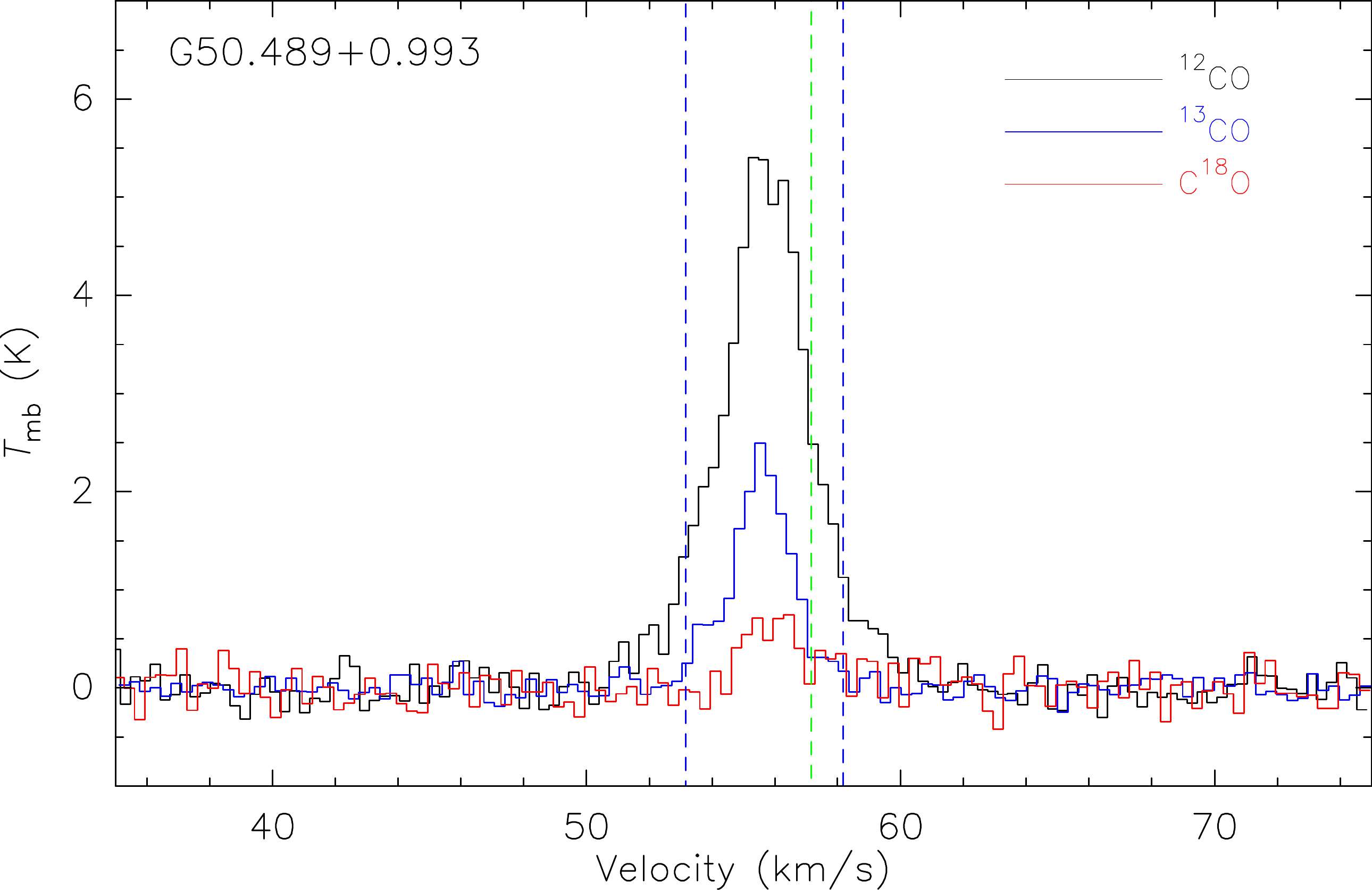}
  \includegraphics[width=6.7cm, angle=0]{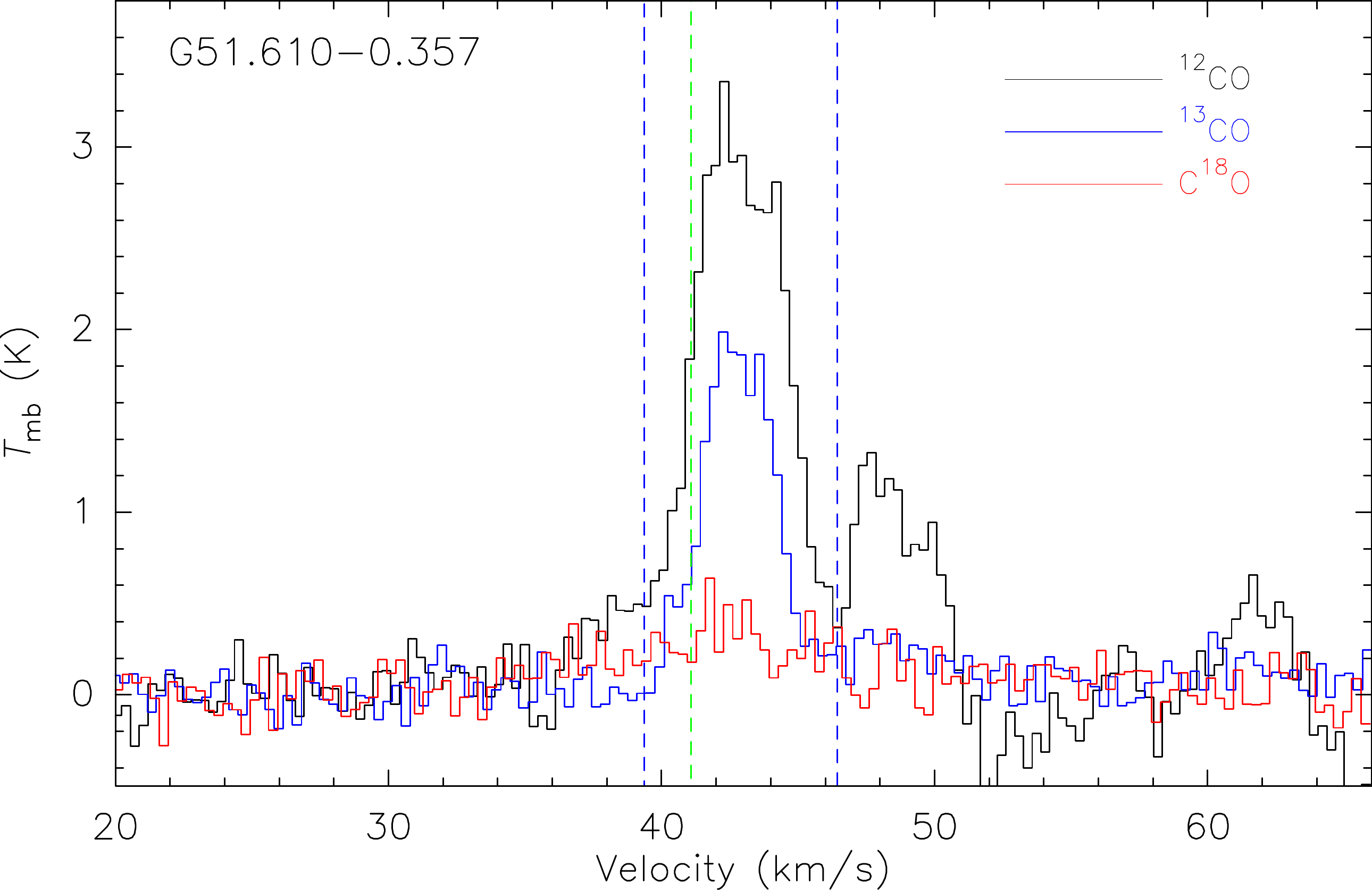}
  \vspace{-4mm}
   \caption{$^{12}$CO $J$=1-0, $^{13}$CO $J$=1-0, and C$^{18}$O $J$=1-0 spectra over the observed region ($>$3$\sigma$) for each bipolar bubbles. The green dashed lines indicate the RRL velocity of each \HII region, while dubble blue and red lines mark the integrated velocity ranges.} 
   \label{Fig3}
   \end{figure*}
  
\subsection{Dust and CO Molecular Emission}
CO observation can reveal whether molecular gas is associated with bipolar bubble along our line of sight. To  check the morphology of molecular gas consistent with each bipolar bubble, we use $^{12}$CO $J$=1-0, $^{13}$CO $J$=1-0, and C$^{18}$O $J$=1-0 lines to trace the molecular gas emision, and use 870 $\mu$m continuum  to trace cool dust emission.

\begin{figure*}[]
   \centering
   \includegraphics[width=30.0cm, angle=0]{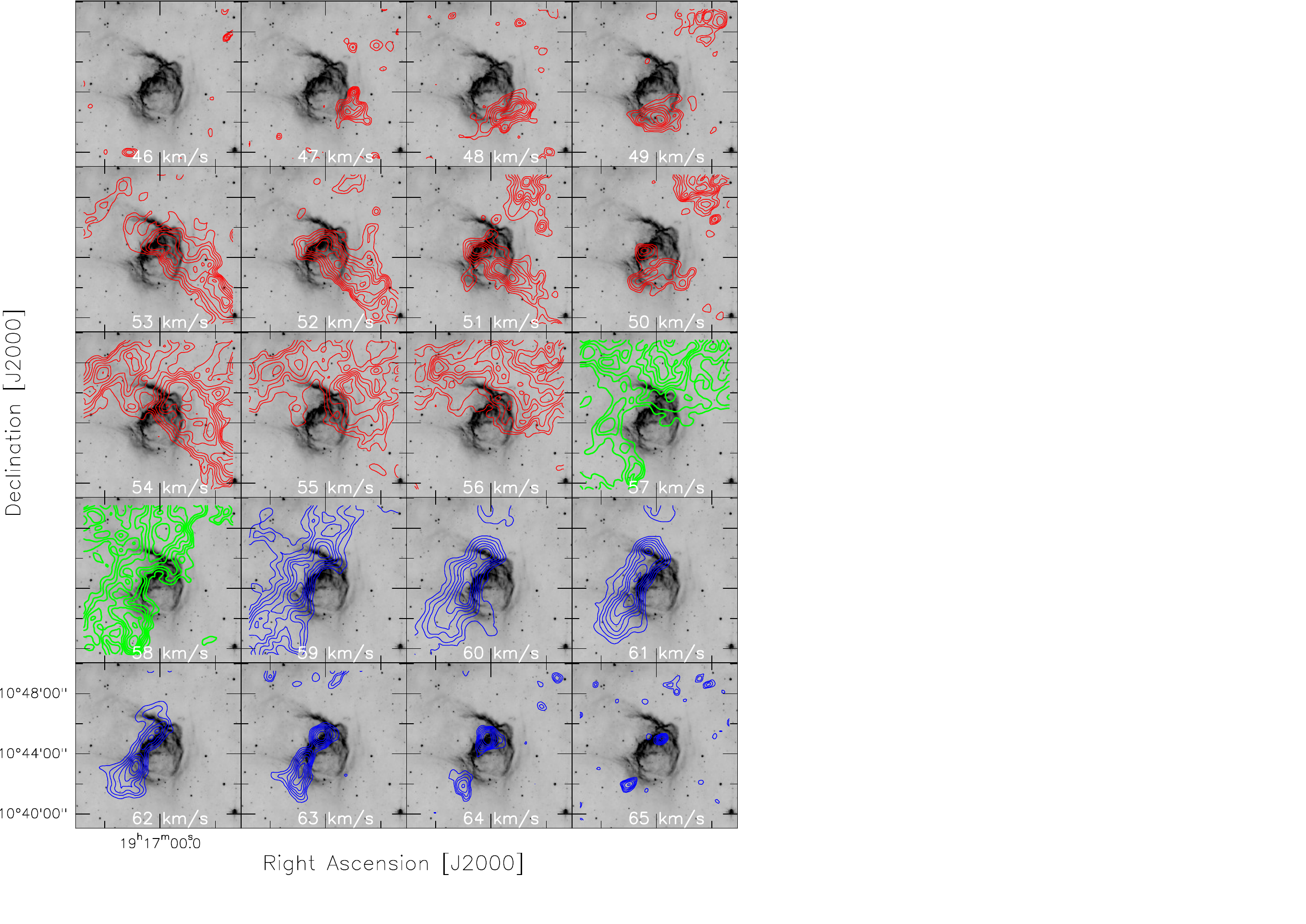}
  \vspace{-18mm}
   \caption{$^{13}$CO $J$=1-0 channel maps of G045.386-0.726 in step of 1 km s$^{-1}$ overlaid on the Spitzer-IRAC 8
$\mu$m emission (grey). Central velocities are indicated in each image. The different gas components are marked by different color contours. } 
   \label{Fig4}
   \end{figure*}

\begin{figure*}[t]
   \centering
   \includegraphics[width=11.0cm, angle=0]{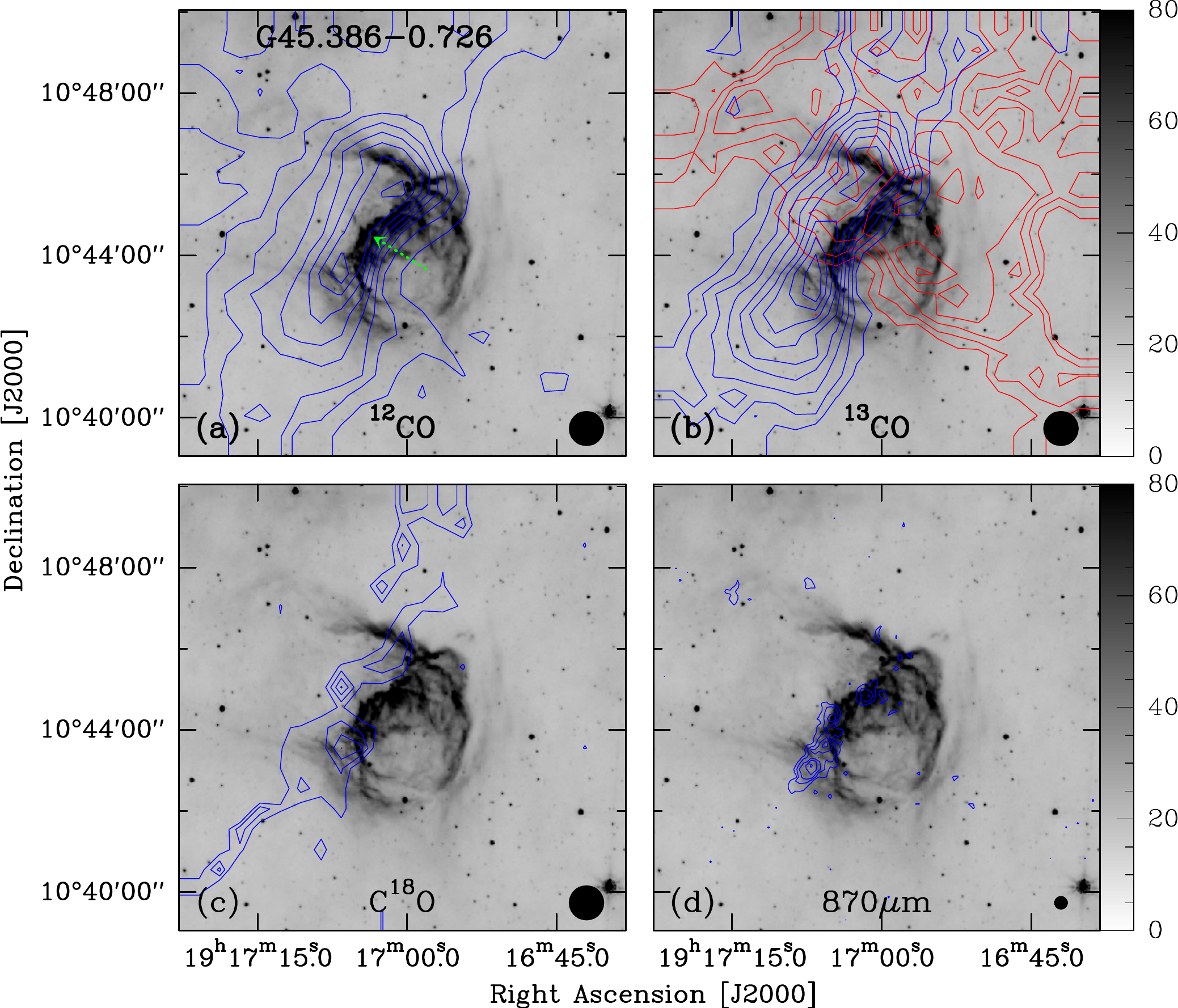}
  \vspace{-4mm}
   \caption{G045.386-0.726. The integrated intensity maps in CO lines  and 870  $\mu$m map are superimposed on the Spitzer  8.0 $\mu$m map (grey scale). (a) panel:   The blue contour levels are from 19.5 (3$\sigma$) to 55.9  by a step of 5.2 K km s$^{-1}$, and the integrated velocity is from 57.0 to 65.5 km s$^{-1}$. (b):  The blue contour levels are from 5.4 (3$\sigma$) to 16.9  by a step of 1.4 K km s$^{-1}$, while the blue contour levels are from 5.4 (3$\sigma$) to 12.6 by a step of 1.8 K km s$^{-1}$. For red contours, the integrated velocity is from 47.0 to 57.0 km s$^{-1}$. (c) panel: The blue contour levels are from 2.1 (3$\sigma$) to 3.8  by a step of 0.6 K km s$^{-1}$. (d) panel: The blue contour levels are from 0.2 (3$\sigma$) to 0.6  by a step of 0.1 Jy/beam. } 
   \label{Fig5}
   \end{figure*}   
 
   \begin{figure}[]
   \centering
   \includegraphics[width=8.2cm, angle=0]{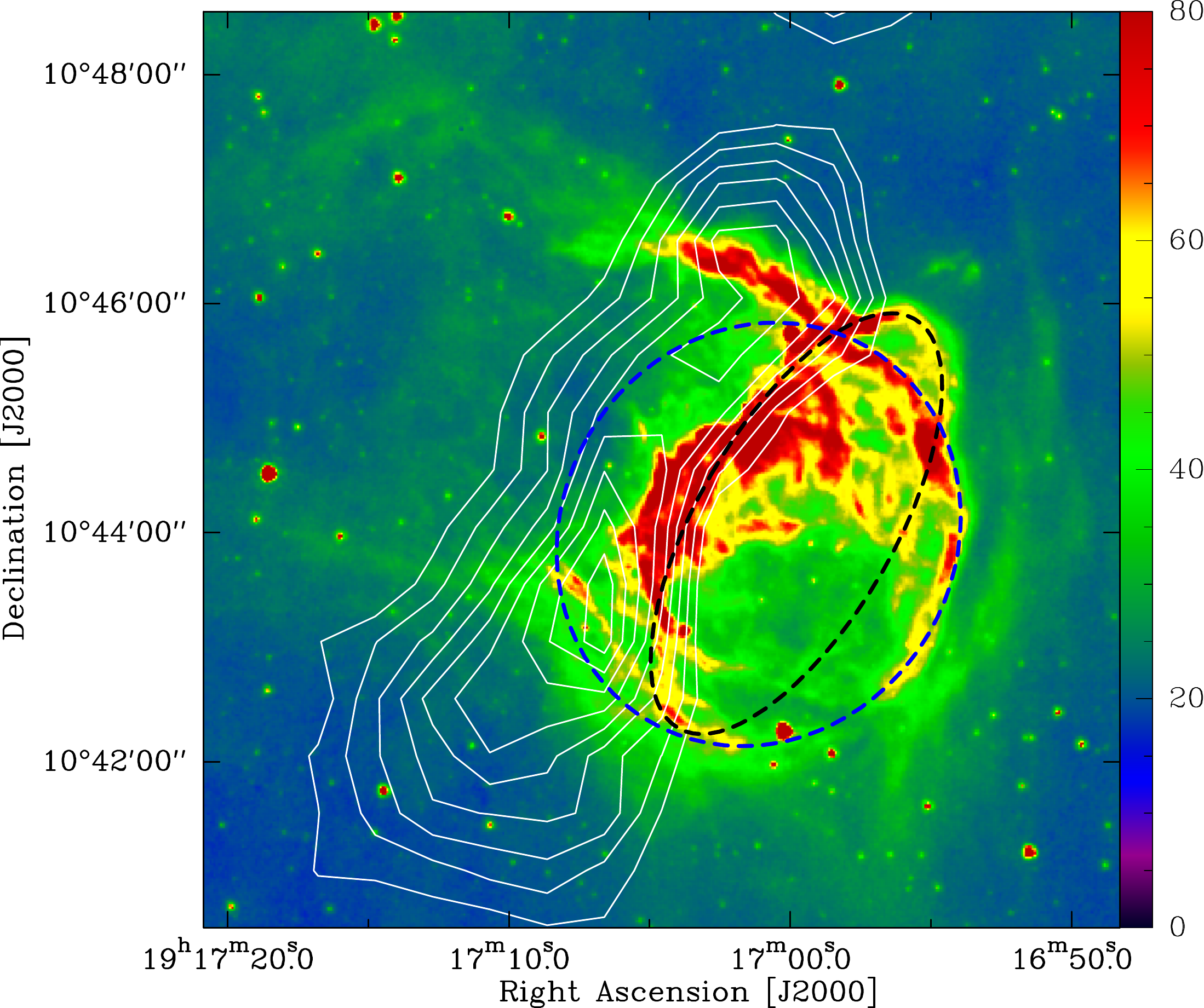}
  \vspace{-4mm}
   \caption{$^{13}$CO $J$=1-0 integrated intensity map (white contours) overlaid on the Spitzer  8.0 $\mu$m map  (color scale). The range of integrated velocity is from 57.0 to 65.0 $\kms$}. The two elliptical dashed circles represent two bubbles.
   \label{Fig6}
   \end{figure}

\subsubsection{G045.386-0.726}
Bipolar bubble G045.386-0.726  is associated with a compact \HII region and bubble N95 \citep{Deharveng2010}. The \HII region has a hydrogen radio recombination line (RRL) velocity of 52.5 $\kms$ \citep{Anderson2011}. From Fig. \ref{Fig3}, we see that there are multiple gas components in the G045.386-0.726 region. Comparing the optically thick $^{12}$CO line, the $^{13}$CO line is more suited to trace relatively dense gas ($\sim$10$^{3}$ cm$^{-3}$). Using channel maps (Fig. \ref{Fig4}) of $^{13}$CO line,  we found two gas components, which may coindcient with G045.386-0.726 in space. One component is located between 47.0 and 58.0 $\kms$ (red+green countours), while the other component is in interval 57.0 to 65.0 $\kms$ (green+blue contours).  The $^{13}$CO emission (green contours) between 57.0 and 58.0 $\kms$ shows that above two components may connect in the velocity and space.   We use these two velocity ranges to make the integrated intensity maps, as shown in Fig. \ref{Fig5}. The $^{12}$CO, $^{13}$CO, and C$^{18}$O emission in interval 57.0 to 65.0 $\kms$,  which is correlated with the 870 $\mu$m emission,  show a filamentary structure from  northwest to southeast. The filament displays an arc-like structure with a steep integrated intensity gradient, whose direction (green arrow) is perpendicular to the filament. The arc-like structure just encloses the bipolar bubble G045.386-0.726.  Moreover, the CO emission in interval 47.0 to 58 $\kms$, which exhibits a diffuse structure, is distribute  mainly over the northwest of G045.386-0.726.  The RRL velocity of \HII region associated with the bipolar bubble G045.386-0.726 is 52.5 $\kms$, indicating that the diffuse CO emission (red contours) is related to bipolar bubble G045.386-0.726, while the filament may be associated with the other bubble. We use two bubbles to fit the Spitzer  8.0 $\mu$m image, as shown in Fig. \ref{Fig6}. Hence, we conclude that these two bubbles with different distances overlap along our sight of line, and thus making it look like a bipolar bubble. Therefore, we will not further analyse  G045.386-0.726 in the rest of the text, except for the discussion section.

  \begin{figure*}[t!]
   \centering
   \includegraphics[width=11.0cm, angle=0]{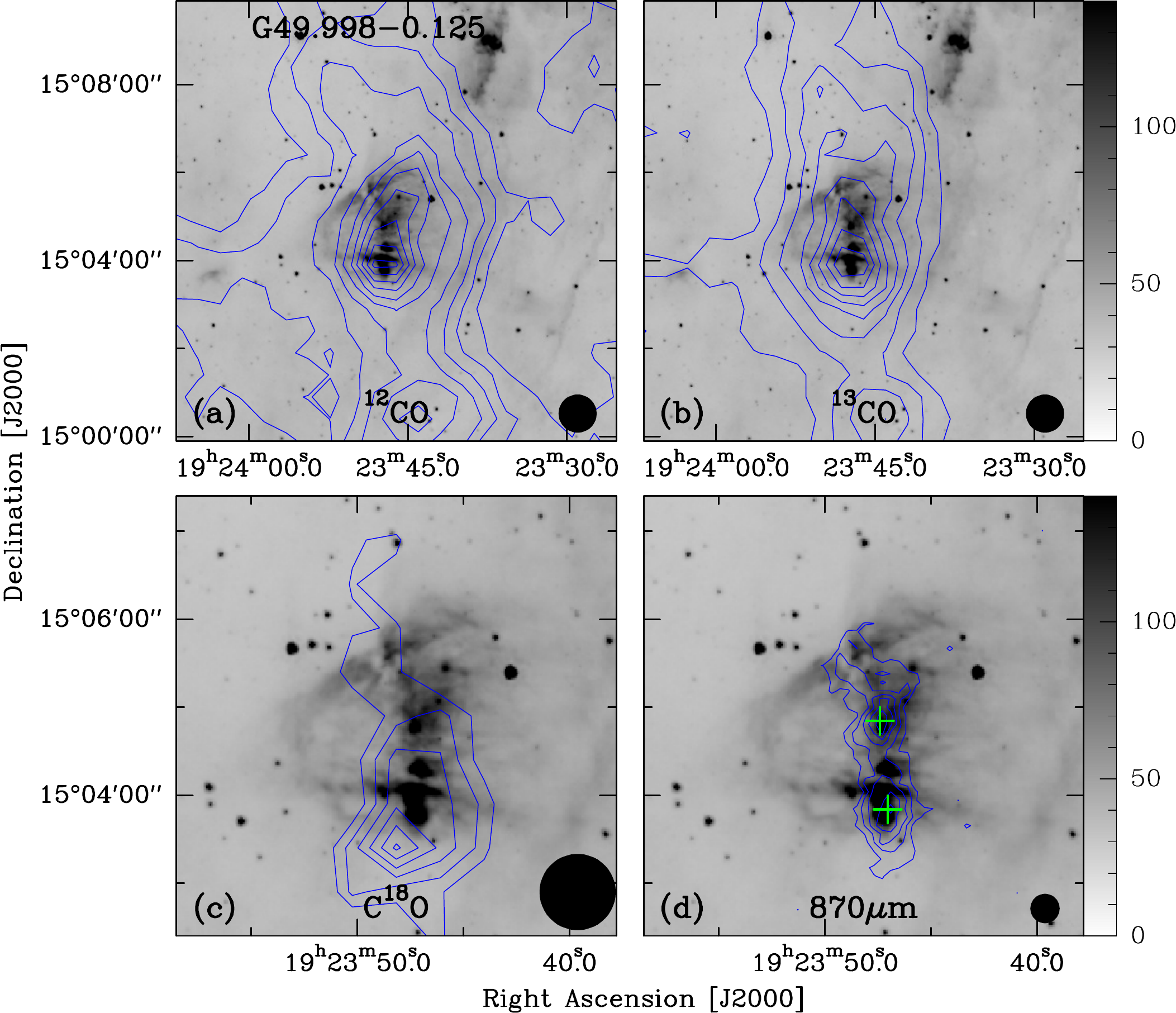}
   \vspace{-4mm}
   \caption{G049.998-0.125. The integrated intensity maps in CO lines  and 870  $\mu$m map are superimposed on the Spitzer  8.0 $\mu$m map (grey scale). (a) panel: The blue contour levels are from 12.0 (3$\sigma$) to 48.0  by a step of 4.0 K km s$^{-1}$. (b) panel: The blue contour levels are from 4.5 (3$\sigma$) to 23.4  by a step of 2.7 K km s$^{-1}$. (c) panel: The blue contour levels are from 2.4 (3$\sigma$) to 4.8  by a step of 0.5 K km s$^{-1}$. (d) panel: The blue contour levels are from 0.2 (3$\sigma$) to 0.8  by a step of 0.1 Jy/beam. The pluses mark the position of the dust clumps.} 
   \label{Fig7}
   \end{figure*}

\subsubsection{G049.998-0.125}
Bipolar bubble G049.998-0.125 is correlated with a compact \HII region. \citet{Anderson2011} gave that this \HII region has two RRL velocities, which are 38.5 $\kms$ and 73.2 $\kms$. Through the channel maps of $^{13}$CO line, we inspect gas component for our studied region, and find that the component in interval 43.2 to 50.0 $\kms$ is associated with G049.998-0.125 in morphology. The two RRL velocities for G049.998-0.125 are not located between 43.2 and 50.0 $\kms$. Using the velocity range of 43.2 to 50.0 $\kms$, we made the integrated intensity maps (Fig. \ref{Fig7}). The $^{12}$CO, $^{13}$CO, and C$^{18}$O emission show a filamentary structure from north to south, but our observations did not resolve the bipolar structure of G049.998-0.125. Comparing the CO emission,  870 $\mu$m emission also displays the filamentary structure with two dense clumps.  Based on the ATLASGAL clump survey \citep{Csengeri2014}, one clump is G049.9996-0.1296, the other clump is G049.9842-0.1364. The masses of the ATLASGAL clumps were determined using the relation
\begin{equation} \mathit{M_{\rm clump}}=\frac{S_{870}D^{2}}{\kappa_{870}B_{870}(T_{d})}
\end{equation}
where $S_{870}$ is the flux density at the frequency 870 $\mu$m and $D$ is the distance to the clumps. 
$\kappa_{870}$ is the dust opacity, which is adopted as 1.0 cm$^{2}$ g$^{-1}$ \citep{Ossenkopf1994}. Here the ratio of gas to dust was taken as 100. $B_{870}(T_{\rm d})$ is the Planck function for the dust temperature $T_{d}$ and frequency 870 $\mu$m. \citet{Deharveng2015} indicated that the mean temperature of the clumps adjacent to the ionized region is about 20 K. Here, we also used a dust temperature of about 20 K for each ATLASGAL clump. The obtained parameters are listed in Table \ref{clumps}. To determine whether the two ATLASGAL clumps have sufficient mass to form massive stars, we must consider their sizes and masses. According to \citet{Kauffmann}, if the clump mass is $M(r)\geq580M_{\odot}(r/pc)^{1.33}$, then they can potentially form massive stars. We find that both the two ATLASGAL clumps lie above  the threshold, indicating that the ATLASGAL clumps are dense and massive enough to potentially form massive stars. Furthermore, the filament in 870 $\mu$m is perpendicular to G049.998-0.125, and divides G049.998-0.125 into two lobes.  The systemic velocity of the filament is 47.0 km s$^{-1}$, which is measured from Fig. \ref{Fig3}.  Based on the Bayesian Distance Calculator \citep{Reid2016},  we obtain that the distance of the filament and G049.998-0.125 is 4.7 kpc.

\subsubsection{G050.489+0.993}
\begin{figure*}[t]
   \centering
   \includegraphics[width=11.0cm, angle=0]{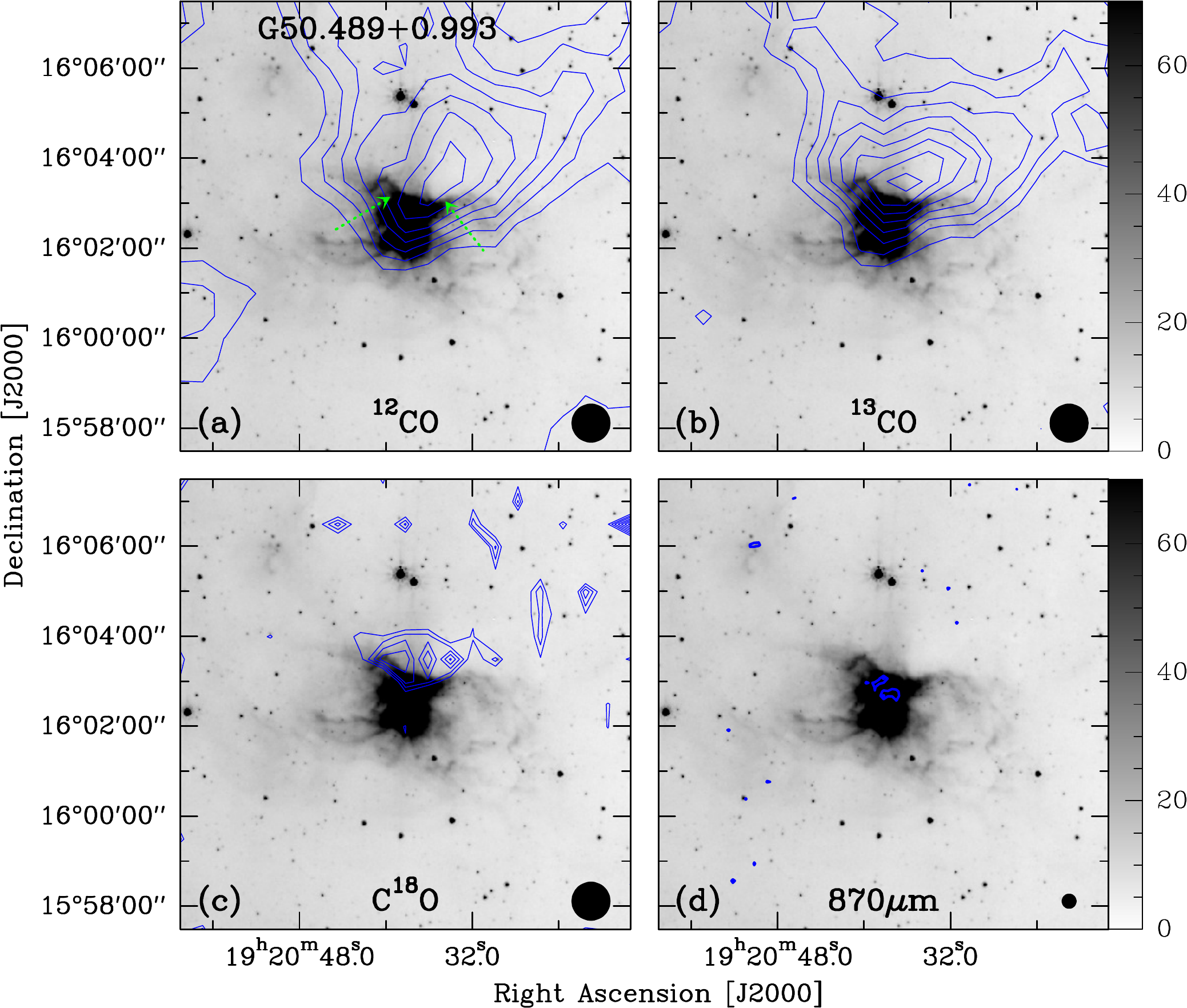}
   \vspace{-4mm}
   \caption{G050.489+0.993. The integrated intensity maps in CO lines  and 870  $\mu$m map are superimposed on the Spitzer  8.0 $\mu$m map (grey scale)(a) panel: The blue contour levels are from 7.5 (3$\sigma$) to 37.5  by a step of 5.0 K km s$^{-1}$. (b) panel: The blue contour levels are from 1.8 (3$\sigma$) to 12.3  by a step of 1.5 K km s$^{-1}$. (c) panel: The blue contour levels are from 0.9 (3$\sigma$) to 1.3  by a step of 0.1 K km s$^{-1}$. (d) panel: The blue contour level is 0.2 Jy/beam (3$\sigma$).} 
   \label{Fig8}
   \end{figure*}

G050.489+0.993  is a textbook example of bipolar bubble. An \HII region is located on the center of G050.489+0.993, whose RRL velocity is 57.1 $\kms$. From Fig. \ref{Fig3}, we note that there is a single component in this region. The  RRL velocity of G045.386-0.726 is just located between 53.1 and 58.1 $\kms$. Hence,  the molecular gas in interval 53.1 to 58.1 $\kms$ is associated with G050.489+0.993. We made the integrated intensity maps, which is shown in Fig. \ref{Fig8}. Both $^{12}$CO and $^{13}$CO emission in  Fig. \ref{Fig8} displays a large clump, while the C$^{18}$O and  870 $\mu$m emission are relatively weak. Moreover, the $^{13}$CO  emission of  the clump  presents a triangle-like shape, and has an integrated intensity gradient toward the direction of two lobes of G050.489+0.993, which are marked in two green arrows. It suggests that the shocks from G050.489+0.993 have expanded into the clump, and have compressed it.  The clump consistent with IRDC G38.95-0.47 also shows the same  shape \citep{Xu2013}, but it is created by different H {\footnotesize II} regions G38.91-0.44 and G39.30-1.04.

\begin{table*}[t]
\small
\tabcolsep 2.9mm\caption{Properties clumps around the bipolar bubble G049.998-0.125\label{clumps}}
\begin{tabular}{@{}lcccccccccccc@{}}
\tableline
Name &  R.A. & Decl.  & $\rm \theta_{maj}\times\rm \theta_{min}$ & $R_{\rm eff}^{\rm a}$ & S &  M($\rm H_{2}$)\\
  & (J2000)  &  (J2000)   & $^{\prime\prime}\times^{\prime\prime}$ & (pc)& (Jy) & M$_{\odot}$\\
\tableline
G049.9996-0.1296& 19$^{\rm h}$23$^{\rm m}$47.40$^{\rm s}$ &  15$^{\rm \circ}$04$^{\rm \prime}$50.8$^{\rm \prime\prime}$ &  34$\times$19 & 0.29&  1.4  &   314.1\\
G049.9842-0.1364 & 19$^{\rm h}$23$^{\rm m}$47.04$^{\rm s}$ &  15$^{\rm \circ}$03$^{\rm \prime}$50.4$^{\rm \prime\prime}$ &  42$\times$23 & 0.35 &  2.0  &  448.8\\
\tableline
\end{tabular}
  \vspace{-4mm}
\tablenotetext{a}{The effective radius of the clumps were calculated as $R_{\rm eff}=\sqrt{\theta_{\rm maj}/2\times\theta_{\rm min}/2}$.}
\end{table*} 

\subsubsection{G051.610-0.357} 
\begin{figure*}[h!]
   \centering
   \includegraphics[width=11.0cm, angle=0]{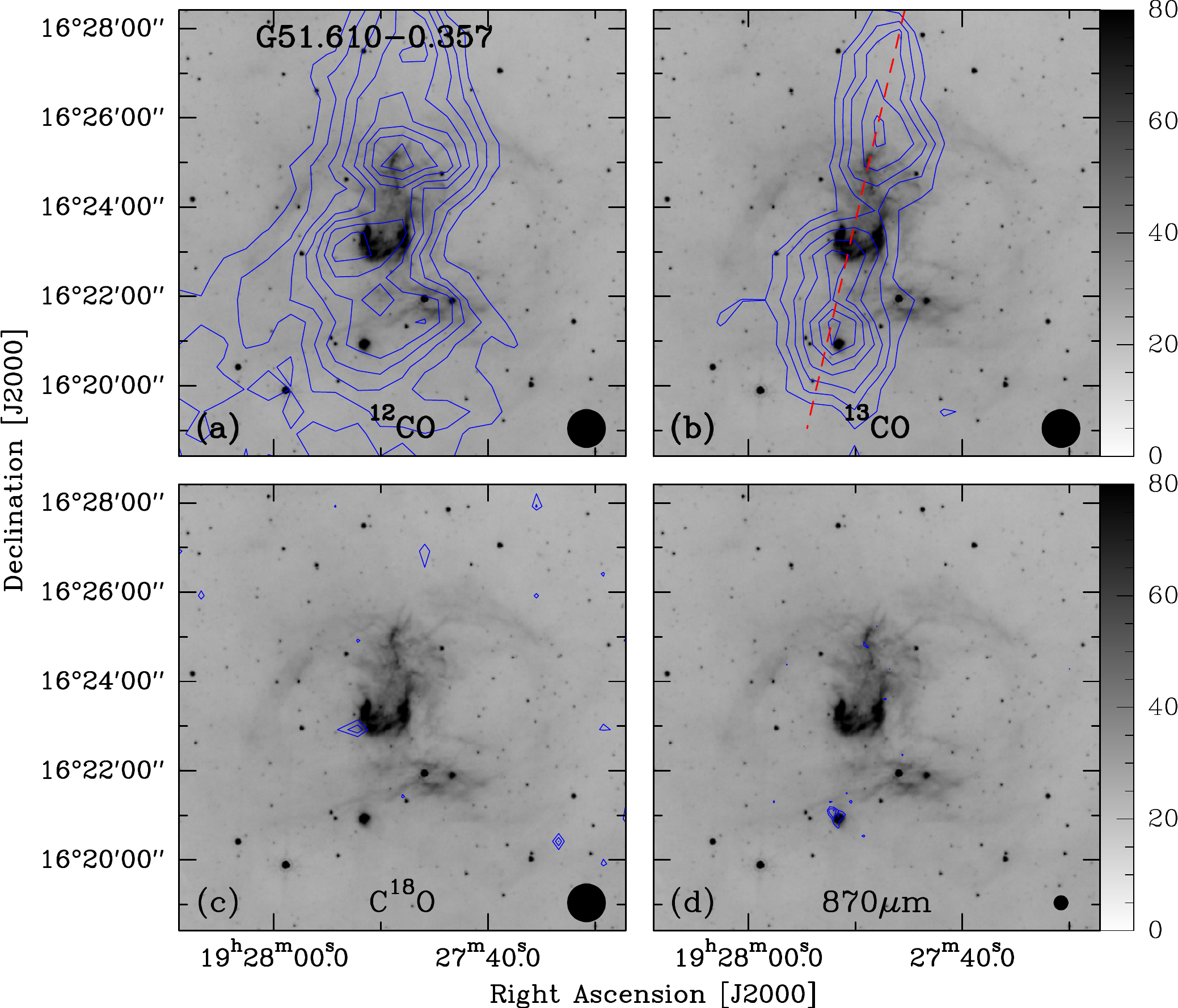}
   \vspace{-4mm}
   \caption{G051.610-0.357. The integrated intensity maps in CO lines  and 870  $\mu$m map are superimposed on the Spitzer  8.0 $\mu$m map (grey scale)(a) panel: The blue contour levels are from 6.5 (3$\sigma$) to 22.1  by a step of 2.6 K km s$^{-1}$. (b) panel: The blue contour levels are from 1.5 (3$\sigma$) to 3.3  by a step of 0.3 K km s$^{-1}$. } 
   \label{Fig9}
   \end{figure*}

Bipolar bubble G045.386-0.726  is also associated with a compact \HII region. This \HII region has a hydrogen RRL velocity of 41.1 $\kms$.  From Fig. \ref{Fig3}, we see that $^{12}$CO spectrum shows multiple components, while $^{13}$CO spectrum exhibits a single component, which is located between 39.1 and 46.1 $\kms$. The integrated intensity maps are shown in Fig. \ref{Fig9}. The $^{12}$CO and $^{13}$CO emission show a filamentary structure  from northwest to southeast.  The filament is just perpendicular to G049.998-0.125, and divides G049.998-0.125 into two lobes. Possiblely the C$^{18}$O and  870 $\mu$m emission is weaker in this region, here we only detect a little emisson for C$^{18}$O and  870 $\mu$m.
  
\subsubsection{Physical parameters} 

To estimate the column density and mass of the molecular gas around the three bipolar bubbles, we used the optical thin $^{13}$CO $J$=1-0 emission.  Assuming local thermodynamical equilibrium (LTE), the column density was estimated via  \citep{Garden}
\begin{equation}
\mathit{N(\rm ^{13}CO)}=4.71\times10^{13}\frac{T_{\rm ex}+0.88}{\rm exp(-5.29/T_{\rm ex})}\int T_{\rm mb}dv ~\rm cm^{-2},
\end{equation}
where  $T_{\rm ex}$ is the mean excitation temperature of the molecular gas, $T_{\rm mb}$ is the corrected main-beam temperature of $^{13}$CO $J$=1-0, and $dv$ is the velocity range.

The $^{12}$CO  emission is optical thick, so we used $^{12}$CO  to estimate $T_{\rm ex}$ via  following the equation  \citep{Garden}
\begin{equation}
\mathit{T_{\rm ex}}=\frac{5.53}{{\ln[1+5.53/(T_{\rm mb}+0.82)]}},
\end{equation}
where $T_{\rm mb}$ is the corrected main-beam brightness temperature of $^{12}$CO.

In addition, we used the relation $N(\rm H_{2})/\it N(\rm ^{13}CO)$ $\approx$
$7\times10^{5}$  \citep{Castets} to estimate the H$_{2}$ column density. The mass of the molecular gas can be determined by
\begin{equation}
\mathit{M_{\rm H_{2}}}=\mu_{g}m(\rm H_{2})\it N(\rm H_{2})\it S,
\end{equation}
where $\mu_{g}$=1.36 is the mean atomic weight of the gas, $m(\rm
H_{2})$ is the mass of a hydrogen molecule, and $S$ is the projected 2D area of the ring. The obtained column density and  mass of the molecular gas are listed in Table \ref{bubble}. From Table \ref{bubble}, we see that the parental molecular gas  in coincidence  with the three bipolar bubbles are the giant molecular cloud.

\section{Discussion}
\label{sect:discu}
From the CO observations, we suggest that there may be two bubbles with different distances overlap along our sight of line, and thus making G045.386-0.726 look like a bipolar bubble. Interestingly, in the velocity interval of  57.0 to 65.0 $\kms$, the $^{12}$CO, $^{13}$CO, and C$^{18}$O emission are correlated with the 870 $\mu$m emission in shape, both of which show a filament. The filament displays an arc-like structure with a steep interated intensity gradient toward G045.386-0.726,  implying that G045.386-0.726 is interacting with the filament. Recently, star formation are found along filaments, so many studies are devoted to filaments. As the suggestion of \citet{Deharveng2015}, it is often forgotten that 2D gas sheets can be mistaken as filaments if we view edge-on.

From Fig. \ref{Fig6}, we also note that the bright 8 $\mu$m emission marked by black dashed lines show a ring-like structure.  Since the ring has an inclination angle with our line of sight, then it appears to display an elliptical structure. The existence of the inclination indicates that the ring is not a 3D spherically symmetrical, but a 2D ring.  The filamentary molecular clouds are usually associated with massive star-forming regions \citep{Andre2014}. Once a massive star forms inside or near a filament, UV radiation and stellar winds will excite the surrounding gas and create an infrared bubble \citep{zhang}. The main emission of the bubble is the bright PAHs. The PAHs molecules are easily excited by the UV radiation from \HII region \citep{Pomares2009}. It means that a bright 8 $\mu$m bubble is created only if there should be some molecular gas adjacent to \HII region, containing the collected gas from the expansion of \HII region. If the bubble is a 2D ring,  the interaction surface between the filament and the ring may be not greater than the thickness of the ring, indicating that the filament in Fig. \ref{Fig4} is a sheet with a thickness of a few parsecs rather than a cylindrical filament.

\begin{figure}[t]
   \centering
   \includegraphics[width=7.0cm, angle=0]{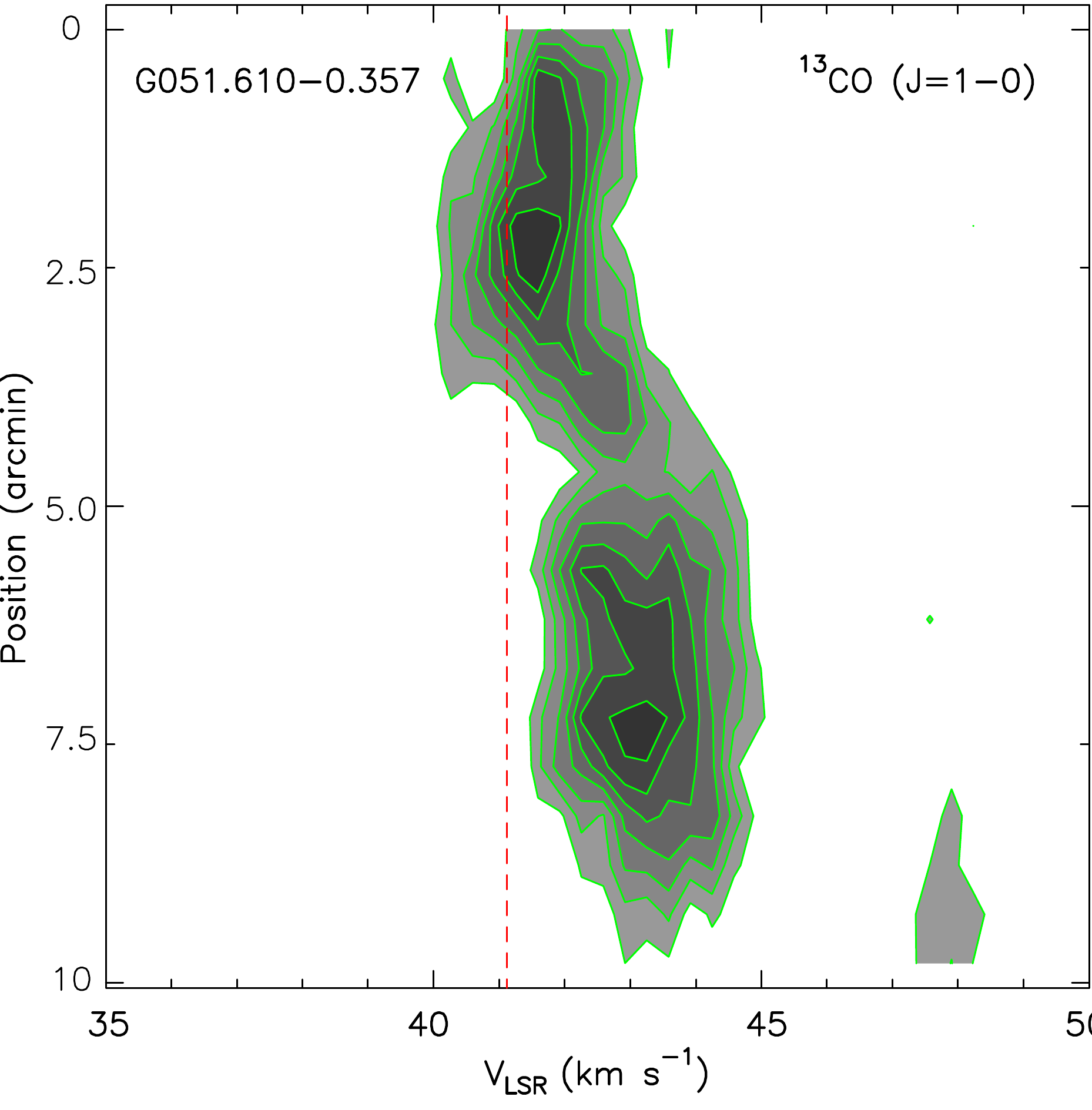}
  \vspace{-4mm}
   \caption{Position-Velocity diagrams of the $^{13}$CO $J$=1-0 emission along the filament associated with G051.610-0.357 (see the long dashed lines from top to bottom in Fig. \ref{Fig9}).  The black dashed line marks the RRL velocity of 41.1 km s$^{-1}$.} 
   \label{Fig10}
   \end{figure}

\begin{figure}[t]
   \centering
   \includegraphics[width=7.5cm, angle=0]{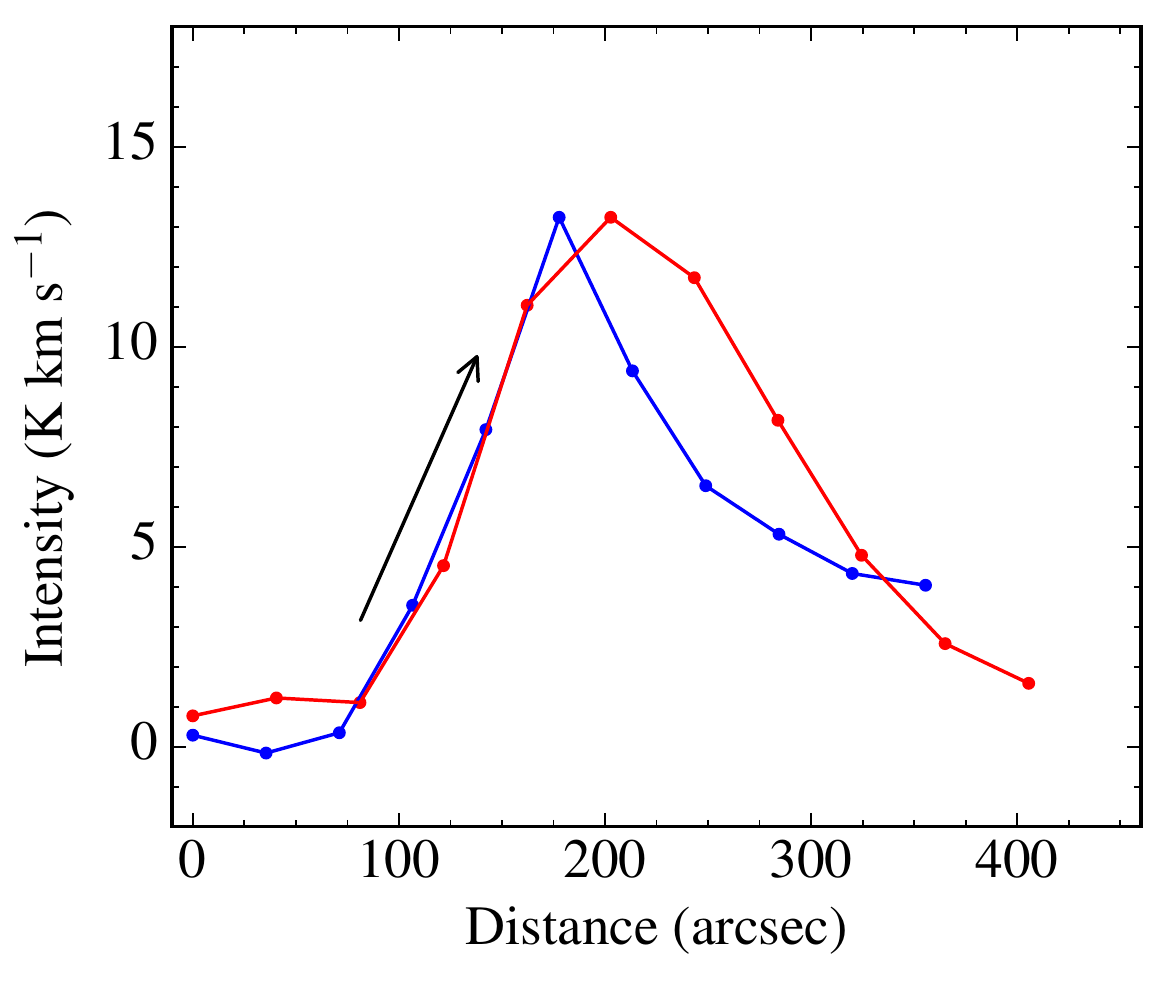}
  \vspace{-4mm}
   \caption{Slice profiles of $^{13}$CO $J$=1-0 emission through the clump in two directions. The cutting paths are shown as the green arrows in Fig. \ref{Fig8}. The black arrow shows a distribution of steep rise.} 
   \label{Fig11}
   \end{figure}

For the three bipolar bubbles (G049.998-0.125, G050.489+0.993, and G051.610-0.357), both the ionized gas and hot dust emission show a single structure. We suggest that each bipolar bubble is created by a single \HII region. Both G049.998-0.125 and G051.610-0.357 are associated with a  filament, respectively. The filament related to G049.998-0.125 contains two ATLASGAL clumps, which are dense and massive enough to potentially form massive stars. Figure \ref{Fig10} shows the position-velocity (PV) diagram constructed from the $^{13}$CO $J$=1-0 emission along the long filament coincident with G051.610-0.357. Since our observations did not resolve the bipolar structure of G049.998-0.125, we only made the PV diagram for the filament associated with G051.610-0.357. From the Fig. \ref{Fig10}, we can discern that the filament is a single coherent object. In  CO emission, both the filaments  are perpendicular to G049.998-0.125 and G051.610-0.357, and divides them into two lobes.  G051.610-0.357 has two symmetrical lobes.  The central position of \HII region is located on the filament for G051.610-0.357. For G049.998-0.125, the left lobe is smaller than the right lobe, it is because the central position of \HII region associated with G049.998-0.125 is far from the filament, through comparing  Fig. \ref{Fig1} and Fig. \ref{Fig7}. The bipolar bubble forms when ionization front breaks through the two opposite faces of the sheet-like cloud simultaneously \citep{Deharveng2010,Deharveng2015}. From our observed results for G049.998-0.125 and G051.610-0.357, they are very similar to  that of \citep{Deharveng2010,Deharveng2015}. We  suggested that the exciting stars of G049.998-0.125 and G051.610-0.357 are formed in the sheet-like structure clouds. The sheet only observed edge-on appears as a filament.  Moreover, the two lobes of G050.489+0.993 is associated with a clump, not a filament. Hence, we can not explain how  bipolar bubble G050.489+0.993 forms. The clump shows a triangle-like  shape with an integrated intensity gradient toward the direction of G050.489+0.993.  Slice profiles of the $^{13}$CO $J$=1-0 emission are shown in Fig. \ref{Fig11} through the clump in two directions. The two cutting directions are shown as the green arrows in Fig. \ref{Fig8}. We found a steep rise at the edges of the clump coincident with G050.489+0.993, as shown in black arrow in Fig. \ref{Fig11}. This steep rise indicates that the two edges of the clump have been compressed by the expansion from the two lobes of  bipolar bubble G050.489+0.993.  We suggest that the shocks from this bipolar bubbles have expanded into the pre-existing clump, and have compressed it. 

\section{CONCLUSIONS}
\label{sect:summary}
We performed the molecular $^{12}$CO $J$=1-0, $^{13}$CO $J$=1-0 and C$^{18}$O $J$=1-0, infrared, and  radio continuum studies towards four bipolar bubbles. The four bipolar bubbles displays the bipolar structure at  the 8.0 $\mu$m. From CO observations, we found that G045.386-0.726 is not a bipolar bubble, but the overlaping two bubbles with different distances along our sight of line. One of the two bubbles is related to a filament. The filament displays an arc-like structure with a steep integrated intensity gradient toward the bubble. We concluded that the filament is a sheet with a thickness of less than a few parsecs rather than a cylindrical filament. Both G049.998-0.125 and G051.610-0.357 are associated with a filament, while G050.489+0.993 is coincident with a clump. The molecular gas in coincidence  with the three bipolar bubbles are the giant molecular cloud. The filaments in  CO emission are perpendicular to G049.998-0.125 and G051.610-0.357, and divide them into two lobes.  We  suggested that the exciting stars of these two bipolar bubbles are formed in a sheet-like structure cloud.  Moreover, the clump associated with G050.489+0.993 shows a triangle-like shape with an integrated intensity gradient toward the two lobes of G050.489+0.993, indicating that bipolar bubble G050.489+0.993 have expanded into the clump.

\acknowledgments
We thank the referee for his/her report which helps to improve the quality of the paper. We are also grateful to the staff at the Qinghai Station of PMO for their assistance during the observations.  This work was supported by the National Natural Science Foundation of China (Grant No. 11363004 and 11403042.)

\nocite{*}
\bibliographystyle{spr-mp-nameyear-cnd}
\bibliography{biblio-u1}

\end{document}